\def\zem{$z_{\rm em}$}
\def\zabs{$z_{\rm abs}$}

\def\l{$\lambda$}

\def\ha{H-$\alpha$}
\def\lya{Ly-$\alpha$}

\def\oiii{[O~{\sc iii}]}

\def\hi{H~{\sc i}}

\def\nhi{\mbox{$\sc N(\sc H~{\sc I})$}}
\def\lognhi{\mbox{$\log \sc N(\sc H~{\sc I})$}}

\def\caii{Ca~{\sc ii}}

\def\cii{C~{\sc ii}}
\def\civ{C~{\sc iv}}
\def\feii{Fe~{\sc ii}}

\def\mgi{Mg~{\sc i}}
\def\mgii{Mg~{\sc ii}}

\def\nii{[N~{\sc II}]}
\def\sii{S~{\sc i}}
\def\siii{Si~{\sc ii}}

\def\znii{Zn~{\sc ii}}
\def\crii{Cr~{\sc ii}}
\def\coii{Co~{\sc ii}}

\def\oii{[O~{\sc ii}]}

\def\sii{S~{\sc ii}}

\documentclass[useAMS,epsfig]{mn2e}

\usepackage{rotating}

\usepackage{graphicx}

\title[Additional SINFONI DLAs Detections]{A SINFONI Integral Field Spectroscopy Survey for Galaxy Counterparts to Damped Lyman-$\alpha$ Systems - III. Three Additional Detections \thanks{Based on observations collected during programmes ESO 83.A-0699 and 85.A-0708 at the European Southern Observatory with SINFONI on the 8.2 m YEPUN telescope operated at the Paranal Observatory, Chile.} }
\author[C\'eline P\'eroux et al.] {C\'eline P\'eroux$^1$\thanks{e-mail:celine.peroux@gmail.com}, Nicolas Bouch\'e$^{2,3,4}$, Varsha P. Kulkarni$^5$,
Donald G. York$^6$,
\newauthor
 \& Giovanni Vladilo$^7$\\
$^1$ Laboratoire d'Astrophysique de Marseille, OAMP, Universit\'e Aix-Marseille \& CNRS,\\
38 rue Fr\'ed\'eric Joliot Curie, 13388 Marseille cedex 13, France  \\
$^2$ Department of Physics, University of California, Santa Barbara, CA 93106, USA; Marie Curie Fellow\\
$^3$ CNRS; Institut de Recherche en Astrophysique et Plan\'etologie [IRAP] de Toulouse,  14 Avenue E. Belin, F-31400 Toulouse, France\\
$^4$ Universit\'e de Toulouse; UPS-OMP; IRAP; F-31400 Toulouse, France\\
$^5$ Dept. of Physics and Astronomy, Univ. of South Carolina, Columbia, SC 29208, USA.\\
$^6$ Dept. of Astronomy and Astrophysics and The Enrico Fermi Institute, University of Chicago, 5640 S. Ellis Ave, Chicago, IL 60637, USA.\\
$^7$ Osservatorio Astronomico di Trieste - INAF, Via Tiepolo 11 34143 Trieste, Italy.
}
\begin{document}

\date{Accepted 2011 October 04. Received 2011 October 04; in original form 2011 January 14}

\pagerange{\pageref{firstpage}--\pageref{lastpage}} \pubyear{2011}

\maketitle

\label{firstpage}

\begin{abstract}
We report three additional SINFONI detections of \ha\ emission line from quasar absorbers, two of which are new identifications. These were targeted among a sample of systems with \lognhi$>$19.0 and metallicities measured from high-resolution spectroscopy. The detected galaxies are at impact parameters ranging from 6 to 12 kpc from the quasar's line-of-sight. We derive star formation rates (SFR) of a few M$_{\odot}$/yr for the two absorbers at \zabs$\sim$1 and SFR=17 M$_{\odot}$/yr for the DLA at \zabs$\sim$2. These three detections are found among a sample of 16 DLAs and sub-DLAs (5 at \zabs$\sim$1 and 7 at \zabs$\sim$2). For the remaining undetected galaxies, we derive flux limits corresponding to SFR$<$0.1--11.0 M$_{\odot}$/yr depending on redshift of the absorber and depth of the data. When combined with previous results from our survey for galaxy counterparts to \hi-selected absorbers, we find a higher probability of detecting systems with higher metallicity as traced by dust-free [Zn/H] metallicity. We also report a higher detection rate with SINFONI for host galaxies at \zabs$\sim$1 than for systems at \zabs$\sim$2. Using the \nii/\ha\ ratio, we can thus compare absorption and emission metallicities in the same high-redshift objects, more than doubling the number of systems for which such measures are possible. 
\end{abstract}

\section{Introduction}

Our understanding of the processes taking place in galaxy formation has considerably advanced in the last years but important issues still need to be addressed. In order to obtain a complete picture, we need to study the
details of processes though which galaxies convert their gas into stars. One way to tackle these phenomena is to relate the neutral hydrogen gas and the stars in individual galaxies. 

The study of the galaxies probed by the absorption that their gas produces in the spectrum of a background quasar has been a prime tool to study the evolution of gas over cosmological time scales (P\'eroux et al. 2003b, P\'eroux et al. 2005, Prochaska et al. 2005, Noterdaeme et al. 2009). In fact, the physical state and chemical abundances of the gas in the absorbing galaxies can be probed back to $\sim$10 billion years ago (e.g. Kulkarni et al. 2007, Kulkarni et al. 2010). Nevertheless, studying the stellar content of all these systems turns out to be rather challenging: the galaxies that produce such absorption might be faint, thus requiring deep observations to detect their stellar/interstellar emission; in addition, they sometimes have small angular separations from the bright background quasars, which makes it difficult to disentangle the light of the galaxy from that of the quasar (Rao et al. 2003, Chen et al. 2003, Rosenberg et al. 2003, Rao et al. 2011).

We use the IFU technique to detect redshifted \ha\ emission from the host of quasar absorbers. This approach has previously led to the detection of 2/3$^{rd}$ of the galaxy hosts in a sample of 21 \mgii-selected absorbers at \zabs$\sim$ 1 (Bouch\'e et al. 2007). More recently, Bouch\'e et al. (2011) have undertook a similar survey at z$\sim$2 but obtain a lower detection rate (four out of twenty \mgii-selected absorbers). Here, we expand our survey for \lognhi$>$19.0 quasar absorbers (P\'eroux et al. 2011a and P\'eroux et al. 2011b) and report additional detections at redshifts z$\sim$1 and z$\sim$2, two of which are new identifications. These three DLAs/sub-DLAs have \nhi\ and hence absorption metallicities known from high-resolution spectroscopy, and are detected among a sample of 16 quasar absorbers. 

The present paper is structured as follows. A summary of the targets selection and observational details are provided in Section 2. 
In the third section, we present the detected DLA/sub-DLAs galaxies. The results on physical properties, including detection rate, optimal target selection criteria, star formation estimates and metallicities are presented in Section 4. Finally, the appendix contains details of the systems which yield no detection. The dynamical properties (see, for instance, P\'eroux et al. 2011b) of these detections will be published in future papers. 
Throughout this paper, we assume a cosmology with H$_0$=71 km/s/Mpc, $\Omega_M$=0.27 and $\Omega_{\rm \Lambda}$=0.73.

\section{Observations}

\begin{table*}
\begin{center}
\caption{Journal of Observations.}
\label{t:JoO}
\begin{tabular}{ccccccccccc}
\hline\hline
Adopted Short 		  &Discovery Quasar Names &V Mag &\zem &\zabs &ESO Period &T$_{\rm exp}$[sec]$\times $N$_{\rm exp}$ &Band &AO$^b$ &seeing["]\\
Quasar Names$^a$&&&&&&&&&\\
\hline
Q0001$-$0159 	&UM 196 	&18.0            &2.810	&2.0950          &P85          &900x4+600x4	        &K &NGS		&0.6\\	
Q0452$-$1640 	&B0449-1645 			&18.0     	&2.679	&1.0072      	&P85         &600x12	        &J &no AO	&1.1\\	
Q1211$+$0902 	&SDSSJ121134.95+090220.9 	&18.5      	&3.292	&2.5841       	&P85                &600x8	        &K &no AO	&0.9\\	
Q1220$-$0040 	&SDSSJ122037.01-004032.4 	&18.6		&1.411	&0.9746		&P83	&600x12		&J	&no AO	&1.0\\
Q1225$+$0035	&SDSS J122556.61+003535.1 	&19.0		&1.226	&0.7731		&P83	&600x4		&J	&no AO	&1.0\\
Q1226$+$1736 	&SDSSJ122607.19+173649.8 	&18.1            &2.925	&2.5576    &P85          &600x4	        &K &NGS		&0.7\\	
Q1234$+$0758	&PKS 1232+082 	&18.4		&2.570	&2.3376		&P83	&600x4		&K	&no AO	&0.9\\
Q1356$-$1101 	&PKS 1354-107 	&19.2            &3.006	&2.5009             &P85                 &600x12	        &K &NGS		&0.5\\	
Q1454$+$1210 	&SDSSJ145418.58+121053.8 	&18.6      	&3.256	&2.2550     &P85   &900x4+600x8	        &K &no AO	&0.7\\	
Q1631$+$1156	&SDSSJ163145.17+115603.3 	&18.5		&1.792	&0.9008		&P83	&600x8		&J	&no AO	&1.1\\
Q2059$-$0528 	&SDSSJ205922.42-052842.7 	&19.3            &2.539	&2.2100             &P85                 &900x8+600x4	        &K &NGS 	&0.9\\	
Q2102$-$3553 	&B2059-360 	&18.6      	&3.090 	&2.5070             &P85                &600x8	        &K &no AO	&1.1\\	
Q2222$-$0946 	&SDSSJ222256.11-094636.2 	&18.3            &2.927	&2.3543             &P85                 &900x8+600x4	        &K &NGS		&0.6\\	
Q2313$-$3704 	&PKS 2311-373 	&18.5      	&2.476	&2.1821             &P85                &600x11	        &K &no AO	&1.0\\	
Q2350$-$0052		&SDSSJ235057.88-005209.8/UM184 		&19.5		&3.023	&2.6147	&P83	&600x8+300x4	&K	&no AO	&0.8\\
Q2352$-$0028 	&SDSSJ235253.51-002850.4 	&18.6            &1.624	&1.0318    &P85   &600x12	        &J &NGS		&0.7\\	
\hline\hline 				       			 	 
\end{tabular}			       			 	 
\end{center}			       			 	 
\begin{minipage}{140mm}
{\bf Note:} \\
{\bf $^a$} the adopted short quasar names are truncated hhmm $\pm$ddmm values for the 2000 coordinates..\\
{\bf $^b$} no AO: no Adaptive Optics, natural seeing. NGS: Adaptive Optics with a Natural Guide Star.\\
\end{minipage}
\end{table*}

We have searched for the H-$\alpha$ emissions in 16 intervening quasars absorbers with \hi\ column densities  19.0 $<$ \lognhi $<$ 20.3, corresponding to the definition of sub-DLAs (P\'eroux et al. 2003a) and \lognhi$>$ 20.3, corresponding to the definition of DLAs (Wolfe et al. 1995). All of the prime target absorbers have well-constrained \nhi\ based on space-based HST observations (Rao, Turnshek \& Nestor 2006). Similarly, the objects have been selected so that the metallicity nearly free from dust-depletion, as traced by zinc [Zn/H]\footnote{We use the usual definition [X/H]=log(X/H)-log(X/H)$_{\odot}$}, has been studied with high-resolution spectroscopy. Because the redshift of the absorbers is accurately known from fits of heavy elements, we have also selected absorbers for which the predicted H-$\alpha$ lines fall in regions free from OH sky lines contamination. From these, we selected a sample of 16 absorbers with absorption redshift 0.7 $<$ \zabs $<$ 2.6 and appropriate RA/Dec for the observing set-up, constituting the "main" targets as listed in Table~\ref{t:JoO}. Five of these targets are at \zabs$\sim$1 and 11 at \zabs$\sim$2, putting the H-$\alpha$ emission line to SINFONI\footnote{See the following url for more information: http://www.eso.org/sci/facilities/paranal/instruments/sinfoni/}  $J$ or $K$-band filters respectively. 

A journal of observations summarising the quasar properties and experimental set-up are also presented in Table~\ref{t:JoO}. The table provides the observing log, the exposure times for each of the objects and the resulting seeing of the combined data. The observations were carried out in Service Mode during two separate observing runs (ESO 83.A-0699 and 85.A-0708) at the European Southern Observatory with SINFONI on the 8.2 m YEPUN telescope. The ESO period P83 ran from April 2009 to September 2009 while the ESO period P85 ran from April 2010 to September 2010. For fields where the quasar itself is bright enough or in which a bright star is available nearby, we have used them as natural guide stars (NGS) for adaptive optics (AO)  in order to improve the spatial resolution. For the fields with no suitable tip-tilt stars, we have used natural seeing (no AO). The resulting seeing ranges from 0.5" to 1.1". We note that most of the observations for the P83 run and some of the P85 run were incomplete and without adaptive optics. The $J$ ($K$) grism provides a spectral resolution of around R$\sim$2000 (4000). The resulting field-of-view (a mosaic of the 8"$\times$8" SINFONI field-of-view resulting in a 10"$\times$10" effective field-of-view) corresponds to a search radius of $\sim$41 kpc at z$\sim$1 and $\sim$43 kpc at z$\sim$2, thus probing rather large impact parameters between the galaxy host and quasar line-of-sight. The data reduction and flux calibration were performed using the same procedures as the ones described in P\'eroux et al. (2011a) with a more recent version of the ESO pipeline (version 1.9.8).

\section{Detected Galaxy Counterparts to Quasar Absorbers}

In this section, we report on the detection of three absorbers in our SINFONI survey. The remaining undetected systems observed during P83 and P85 are described in the appendix. The redshift of the main absorbers targeted is indicated in the title of the sub-section, while additional systems reported in the literature but {\it not optimal} for SINFONI \ha\ detections in terms of wavelength coverage and OH contamination are indicated in brackets. In some cases, \oiii\ was covered and looked for since \oiii\ can be much stronger than \ha\ in some situations. However, we do not detect any \oiii\ emission lines at flux levels comparable to the one listed in Table~\ref{t:results} for \ha. All of our main targets (and some of the non-prime systems) have been observed in absorption at high-resolution and we indicate abundance ratios, \mgii\ equivalent widths, heavy element velocity widths and number of components whenever available. We also list any previous attempts to image some of these fields.

\subsection{Q0452$-$1640, \zabs=1.0072}

\begin{figure*}
\begin{center}
\includegraphics[height=4.5cm, width=4.5cm, angle=0]{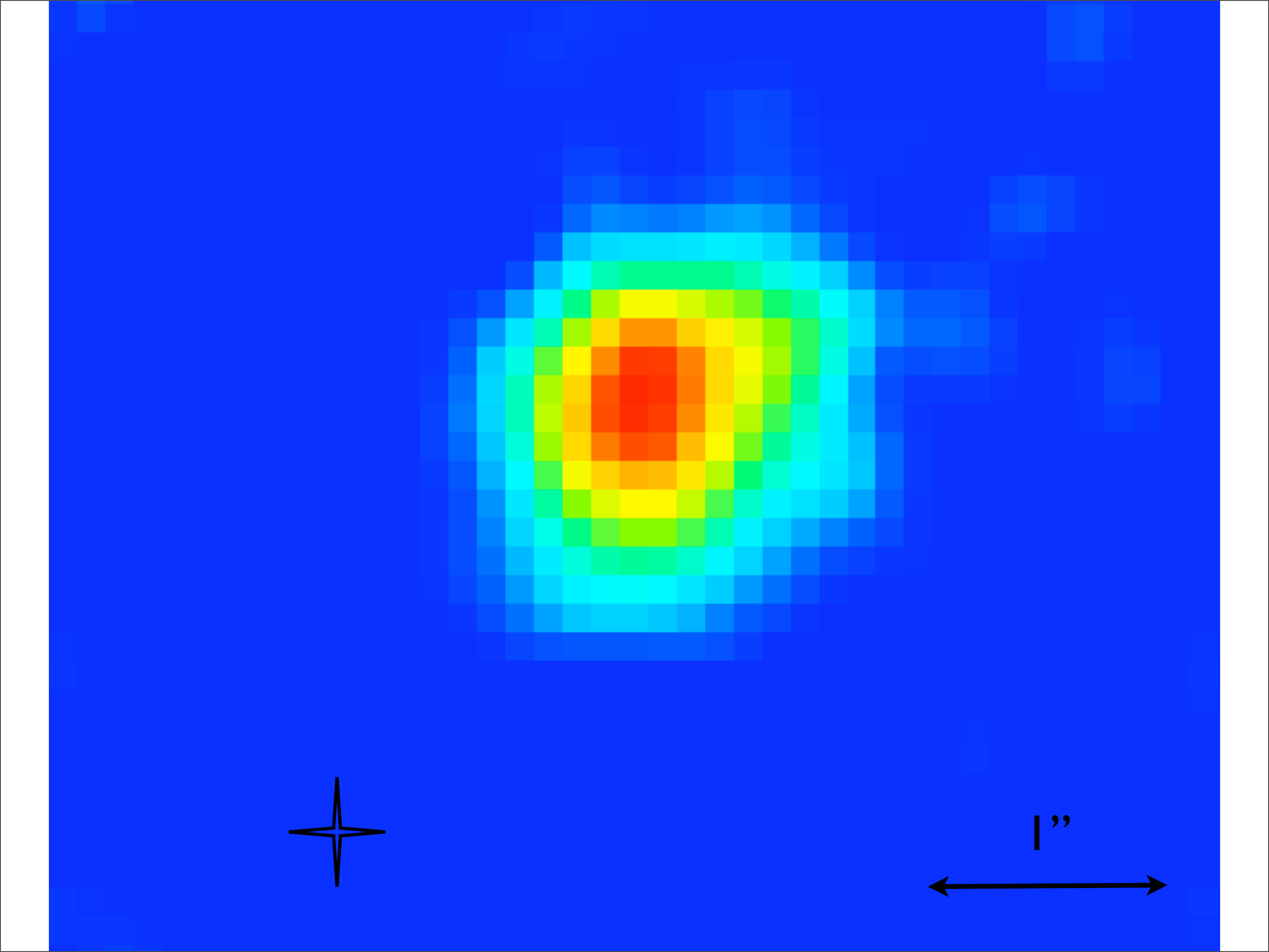}
\includegraphics[height=5cm, width=7.5cm, angle=0]{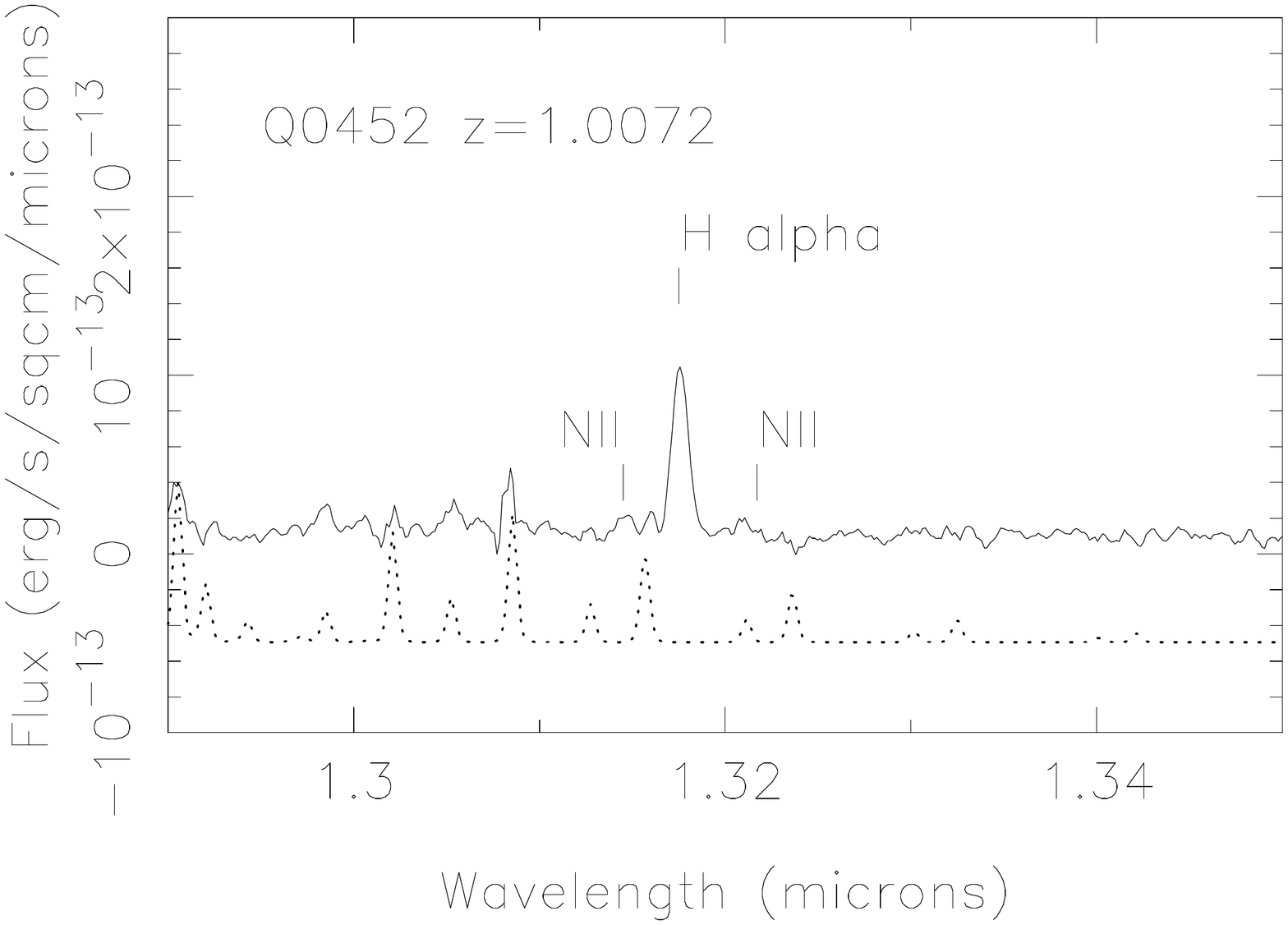}
\caption{{\bf Left Panel:} The H-$\alpha$ emission map of the targeted absorber in the field of Q0452$-$1640  at \zabs=1.0072. At this redshift, 1"=8.0 kpc. The seeing of the stacked image is 1.1". In this, and the following maps, the cross indicates the position of the quasar, north is up and east is to the left. {\bf Right Panel:} The integrated spectrum  over 13 pixels of the corresponding absorber galaxy with the redshifted H-$\alpha$ emission line. In this, and the following figures, the units are in erg/s/cm$^2$/$\mu$. The spectrum is smoothed (5 pixel boxcar). The dotted spectrum at the bottom of the panel is the sky spectrum with arbitrary flux units, scaled for clarity, and indicating the position of the OH sky lines.   }
\label{f:Q0452}
\end{center}
\end{figure*}

This absorber at  \zabs=1.0072 had its \nhi\ column density measured by Rao, Turnshek \& Nestor (2006) to be \lognhi=20.98$^{+0.06}_{-0.07}$. Nestor et al. (2008) reported \znii\ and \crii\ detections in this \caii\ absorber using ISIS (Intermediate dispersion Spectrograph and Imaging System) on WHT (William Herschel Telescope). The quasar was first observed at high-resolution by P\'eroux et al. (2008) who then derived the metallicity of the absorber by fitting 13 components over $\sim$230 km/s to the absorption profile. These authors derive [Mg/H]=$-$0.66$\pm$0.11 from \mgi\ $\lambda$ 2852, [Cr/H]=$-$1.14$\pm$0.09, [Mn/H]=$-$1.54$\pm$0.11, [Fe/H]=$-$1.34$\pm$0.08, [Zn/H]=$-$0.96$\pm$0.08 and [Ti/H]$<-$2.11. The \mgii\ doublet for that system falls in an UVES spectral coverage gap. They also report possible detection of \coii\ in the same system, [Co/H]$=-$0.45. 

This absorber is the primary target of our \ha\ search and is clearly detected in our SINFONI data. Figure~\ref{f:Q0452} shows the \ha\ emission map and the spectrum of this galaxy. Table~\ref{t:results} lists the resulting \ha\ flux, luminosity and SFR estimate. We report F(\nii \l6585)=1.8$\pm$0.5$\times$10$^{-17}$ erg/s/cm$^{2}$, but \sii\ is not detected.

\subsection{Q2222$-$0946, \zabs=2.3541 (2.8647)}

\begin{figure*}
\begin{center}
\includegraphics[height=4.5cm, width=4.5cm, angle=0]{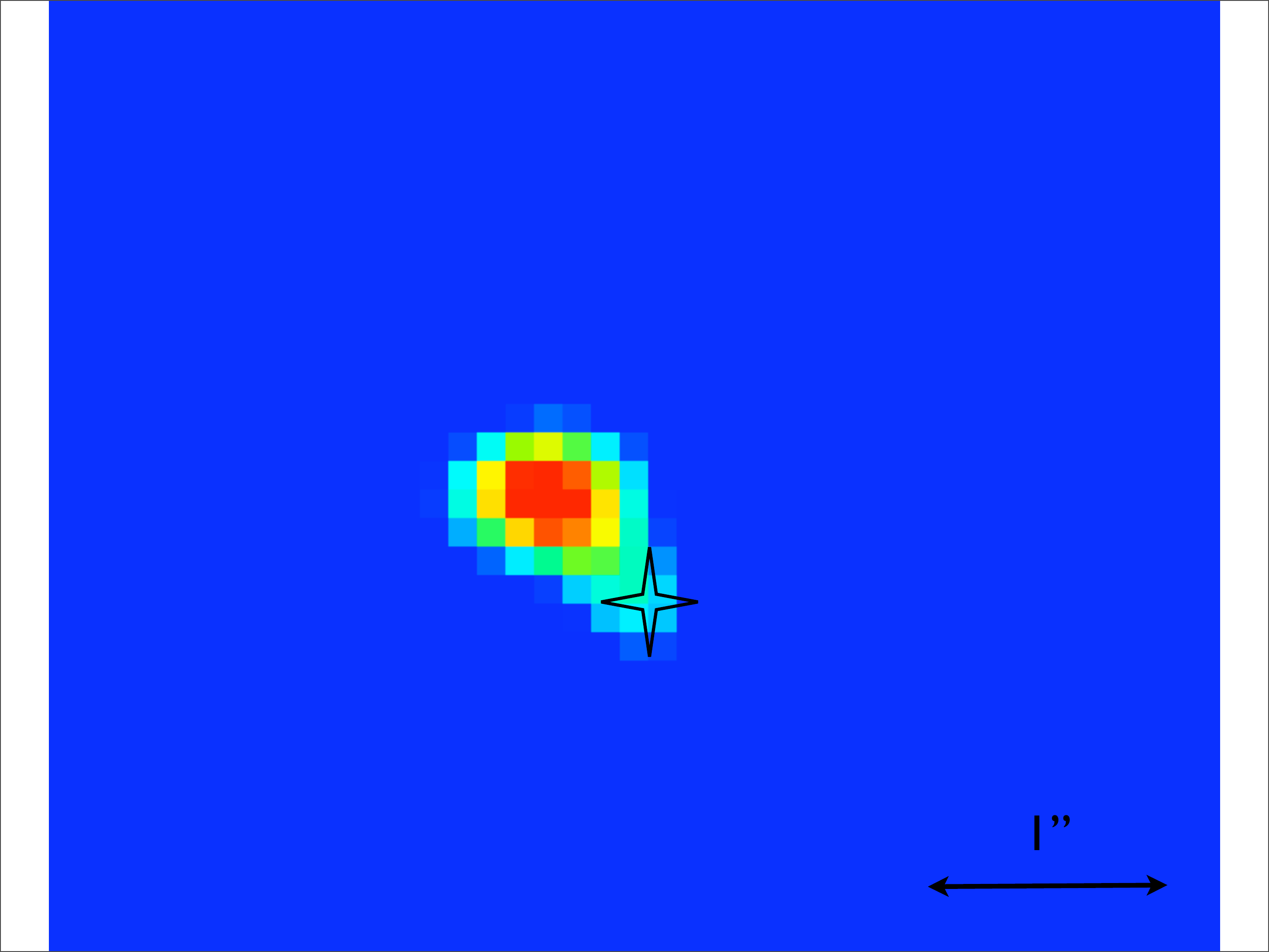}
\includegraphics[height=5cm, width=7.5cm, angle=0]{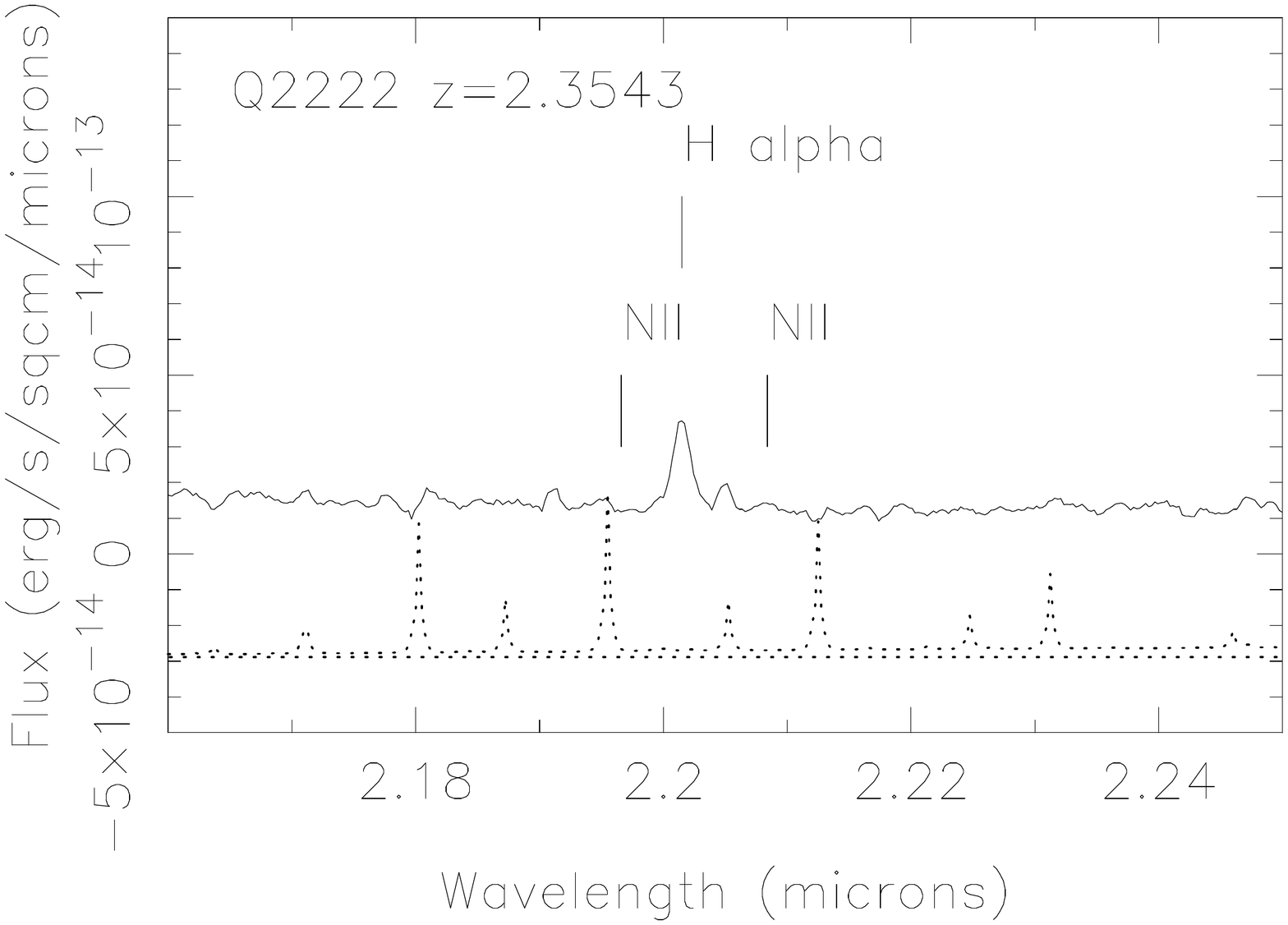}
\caption{{\bf Left Panel:} The H-$\alpha$ emission map of the targeted absorber in the field of Q2222$-$0946  at \zabs=2.3541. At this redshift, 1"=8.3 kpc. The seeing of the stacked image is 0.6". {\bf Right Panel:} The integrated spectrum over 5 pixels of the corresponding absorber galaxy with the redshifted H-$\alpha$ emission line. In that case, the continuum flux level of the galaxy does not go to zero because of the proximity of the quasar.}
\label{f:Q2222}
\end{center}
\end{figure*}

This is a quasar discovered as part of the SDSS survey data release 3 (Schneider et al. 2005). 
Herbert-Fort et al. (2006) have studied the targeted absorber in detail. They report [Si/H]=$-$0.61$\pm$0.20. Prochaska et al. (2007) list \lognhi=20.50$\pm$0.15, [Zn/H]$<-$0.39$\pm$0.11 and [Fe/H]=$-$1.05$\pm$0.02. The \mgii\ doublet is not covered by these data. More recently, Fynbo et al. (2010) have used a triangulation of X-Shooter spectra of that object to both accurately measure the metallicity of the absorber at \zabs=2.3541 fitted by 3 components over $\sim$100 km/s: [Si/H]=$-$0.51$\pm$0.06, [Zn/H]=$-$0.46$\pm$0.07, [Fe/H]=$-$0.99$\pm$0.06, [Ni/H]=$-$0.85$\pm$0.06 and [Mn/H]=$-$1.23$\pm$0.06 and detect \lya\ emission from the galaxy host. They also report EW(\mgii\ 2796)=2.7\AA.

They also detect \ha\ and derive F(\ha)$>$2.5$\times$10$^{-17}$ erg/s/cm$^2$ corresponding to L(\ha)$>$1.1$\times$10$^{42}$ erg/s and SFR$>$10 M$_{\odot}$/yr. They obtain \lya/\ha$>$3.6 and \oiii/\ha=1.6, such ratios are often considered the result of low abundance gas which is hot ($\sim$15K instead of 8K for normal abundances). The \oii\ and H-$\beta$ are located on top of bright sky emission lines leading to an upper limit F(\oii)$<$8$\times$10$^{-17}$ erg/s/cm$^{2}$, i.e. SFR$<$40 M$_{\odot}$/yr. The authors note however that these fluxes are uncertain because the quasar flux in the X-Shooter spectrum seem to be 30\% larger compared to the Sloan calibration and thus they artificially decreased the flux level. In addition, their slit does not actually cover the galaxy itself and the flux calibration is therefore subject to important slit loss. 

We also detect the \ha\ emission in our SINFONI data cube and are able to locate the galaxy on the sky (see Table~\ref{t:detections} for sky position). Figure~\ref{f:Q2222} shows the \ha\ emission map and the spectrum of this galaxy. Table~\ref{t:results} lists the resulting \ha\ flux, luminosity and SFR estimate. We report F(\nii)$<$0.4$\times$10$^{-17}$ erg/s/cm$^{2}$ from a non-detection. From \ha, we derive a SFR estimate of 17.1$\pm$6.0 M$_{\odot}$/yr consistent with findings from Fynbo et al. (2010) who estimated SFR$>$ 10 M$_{\odot}$/yr.

\subsection{Q2352$-$0028, \zabs=1.0318 (0.8730, 1.2468)}

\begin{figure*}
\begin{center}
\includegraphics[height=4.5cm, width=4.5cm, angle=0]{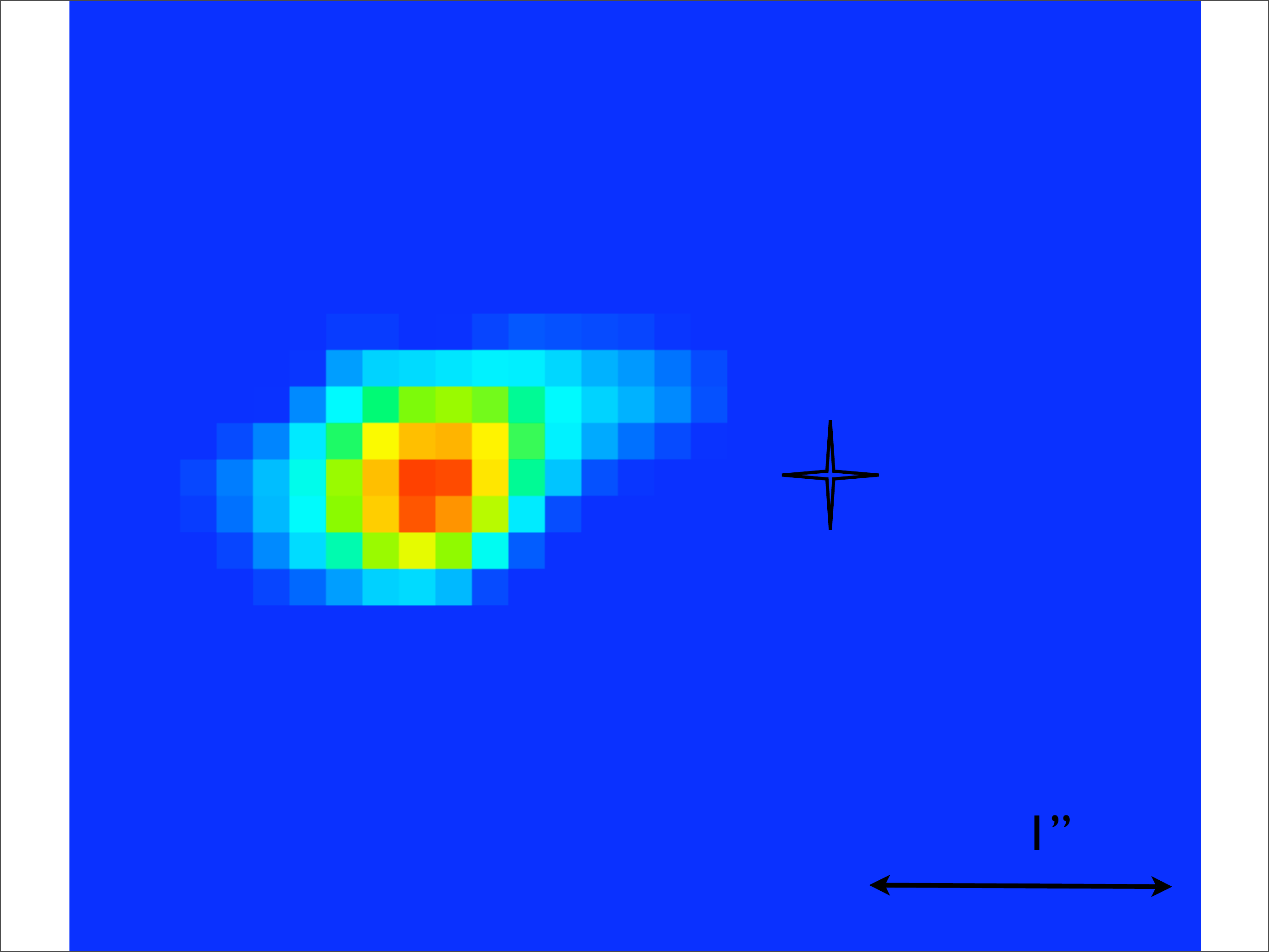}
\includegraphics[height=5cm, width=7.5cm, angle=0]{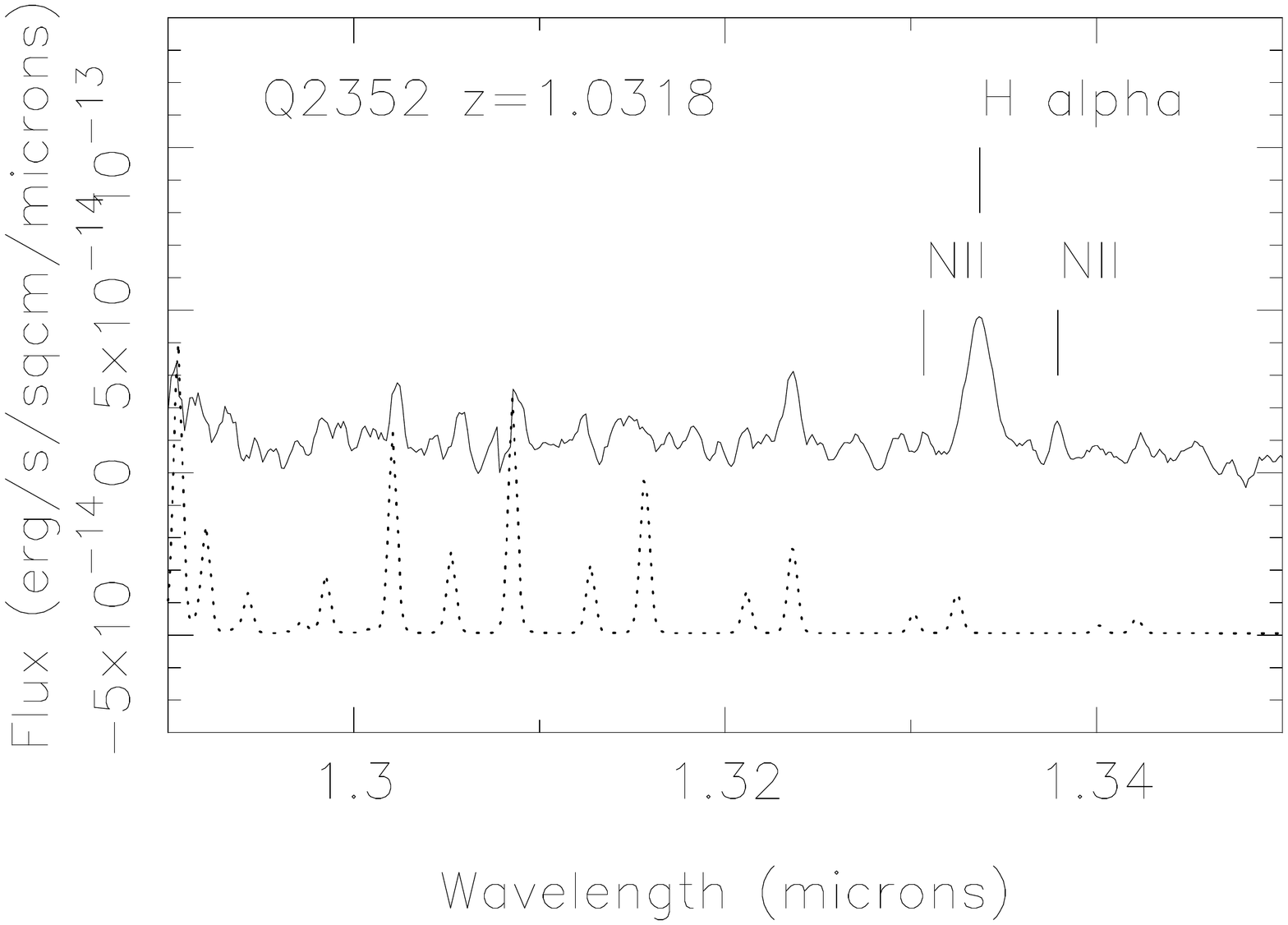}
\caption{{\bf Left Panel:} The H-$\alpha$ emission map of the targeted absorber in the field of Q2352$-$0028  at \zabs=1.0318. At this redshift, 1"=8.3 kpc. The seeing of the stacked image is 0.7". {\bf Right Panel:} The integrated spectrum over 5 pixels of the corresponding absorber galaxy with the redshifted H-$\alpha$ emission line.}
\label{f:Q2352}
\end{center}
\end{figure*}

This quasar was first discovered by Foltz et al. (1989) as part of the Large, Bright Quasar Survey (LBQS). It is also part of the Early Release of the SDSS (Schneider et al. 2002). Rao, Turnshek \& Nestor (2006) list three sub-DLAs with \zabs=0.8730, 1.0318, 1.2468 (see the appendix for details on the non-prime targets).

For the \zabs=1.0318, Rao, Turnshek \& Nestor (2006) measure \lognhi=19.81$^{+0.14}_{-0.11}$. The corresponding EW(\mgii\ 2803)=1.714$\pm$0.110\AA. Meiring et al. (2009) used 11 components to fit the absorber over $\Delta$ v $\sim$ 220 km/s and derive: [Zn/H]$<-$0.51, [Fe/H]=$-$0.37$\pm$0.13 and [Si/H]=$+$0.14$\pm$0.14. 

In our SINFONI observations, we report the detection of the system at \zabs=1.0318. Figure~\ref{f:Q2352} shows the \ha\ emission map and the spectrum of this galaxy. Table~\ref{t:results} lists the resulting \ha\ flux, luminosity and SFR estimate. We report F(\nii)=0.6$\pm$0.3$\times$10$^{-17}$ erg/s/cm$^{2}$, but \sii\ is not detected.

\section{Results}

\subsection{Detection Rate}

\begin{table*}
\begin{center}
\caption{Summary of the detections and upper limits for the 16 main targets from this work. We also recall previous results from the first epoch of our survey (P\'eroux et al. 2011a).}
\label{t:results}
\begin{tabular}{cccccccccc}
\hline\hline
Quasar 		  &\zabs &\lognhi{\bf $^a$} &[Zn/H] {\bf $^a$} &[Fe/H]{\bf $^a$}  &[Si/H] {\bf $^a$} &F(H-$\alpha$){\bf $^b$} &Lum(H-$\alpha$) &SFR  &Ref\\
 		    & &[atoms/cm$^2]$ &&&  &[erg/s/cm$^2]$ &[erg/s]&[M$_{\odot}$/yr]  &\\
\hline
Q0001$-$0159 	&2.0950    		&20.70$\pm$0.10			&$-$0.75		&$-$1.70	&$-$0.84	&$<$4.9$\times$10$^{-17}$	&$<$16.1$\times$10$^{41}$	& $<$ 7.1	&this work	\\
Q0134$+$0051    		     &0.8420 &19.93$^{+0.10}_{-0.15}$  &$<-$0.36		     &$-$0.91	&...		&$<$1.0$\times$10$^{-17}$		&$<$0.3$\times$10$^{41}$	    &$<$0.2 &P\'eroux et al. (2011a) \\
Q0302$-$223			     &1.0094 &20.36$^{+0.11}_{-0.11}$  &$-$0.51		     &$-$1.20	&$-$0.73	&7.7$\pm$2.7$\times$10$^{-17}$			&4.1$\pm$1.4$\times$10$^{41}$		    &1.8$\pm$0.6 	&P\'eroux et al. (2011a)\\
Q0452$-$1640 	&1.0072			&20.98$^{+0.06}_{-0.07}$	&$-$0.96      	&$-$1.34	&...			&14.5$\pm$4.3$\times$10$^{-17}$	&8.0$\pm$2.4$\times$10$^{41}$	&3.5$\pm$1.0	&this work	\\
Q1009$-$0026	   		     &0.8426 &20.20 $^{+0.05}_{-0.06}$ &$<-$0.98		     &$-$1.28	&...		&$<$1.0$\times$10$^{-17}$		 &$<$0.3$\times$10$^{41}$	    &$<$0.2	&P\'eroux et al. (2011a)\\
--			     		     &0.8866 &19.48  $^{+0.05}_{-0.06}$ &$+$0.25		     &$-$0.37	&$<-$0.02	&17.1$\pm$6.0$\times$10$^{-17}$ 		&6.6$\pm$2.3$\times$10$^{41}$		    &2.9$\pm$1.0	 &P\'eroux et al. (2011a)\\
Q1211$+$0902 	&2.5841       		&21.40$\pm$0.10			&$-$1.09		&$-$1.65	&$-$1.05	&$<$2.3$\times$10$^{-17}$	&$<$12.6$\times$10$^{41}$	& $<$ 5.6	&this work	\\
Q1220$-$0040	&0.9746			&20.20$^{+0.05}_{-0.09}$	&$<-$1.14	&$-$1.33	&...				&$<$0.7$\times$10$^{-17}$	&$<$0.3$\times$10$^{41}$	&$<$0.1	&this work\\
Q1225$+$0035	&0.7731			&21.38$^{+0.11}_{-0.12}$	&$-$0.78	&$-$1.19	&...				&$<$5.3$\times$10$^{-17}$	&$<$1.5$\times$10$^{41}$	&$<$0.7	&this work	\\
Q1226$+$1736	&2.5576    		&19.32$\pm$0.15			&$<-$0.36	&$-$1.84	&$-$1.59	&$<$2.9$\times$10$^{-17}$	&$<$15.7$\times$10$^{41}$	& $<$ 7.0	&this work	\\
Q1228$-$113			     &2.1929 &20.60$^{+0.60}_{-0.60}$   &$<-$0.22       	     &...		&...		&$<$0.5$\times$10$^{-17}$		 &$<$2.0$\times$10$^{41}$	    &$<$0.9 &P\'eroux et al. (2011a)\\
Q1234$+$0758	&2.3376			&20.80$\pm$0.10			&$-$0.86		&$-$1.59	&$-$1.18	&$<$1.2$\times$10$^{-17}$	&$<$5.9$\times$10$^{41}$	&$<$2.4	&this work	\\
Q1323$-$0021			     &0.7160 &20.21$^{+0.21}_{-0.18}$   &$+$0.61		     &$-$0.51	&...		&$<$1.0$\times$10$^{-17}$		 &$<$0.2$\times$10$^{41}$	    &$<$0.1&P\'eroux et al. (2011a)\\
Q1356$-$1101 	&2.5009			&20.40$\pm$0.49			&$<-$1.32       	&$-$1.25			&...	&$<$2.3$\times$10$^{-17}$	&$<$12.1$\times$10$^{41}$	& $<$ 5.4	&this work	\\  
Q1454$+$1210 	&2.2550 			&20.30$\pm$0.15			&$-$1.12		&$-$1.47	&$>-$0.40	&$<$1.2$\times$10$^{-17}$	&$<$4.8$\times$10$^{41}$	& $<$ 2.1	&this work	\\
Q1631$+$1156	&0.9008			&19.70$^{+0.03}_{-0.04}$	&$-$0.18		&$-$1.06	&...				&$<$1.5$\times$10$^{-17}$	&$<$0.6$\times$10$^{41}$	&$<$0.3	&this work	\\
Q2059$-$0528 	&2.2100             	&20.80$\pm$0.20			&$-$0.53		&$-$1.30 &$-$1.00	&$<$0.8$\times$10$^{-17}$	&$<$3.2$\times$10$^{41}$	& $<$ 1.4	&this work	\\
Q2102$-$3553 	&2.5070             	&20.21$\pm$0.10			&$<-$0.44	&$-$2.24	&$-$1.04  &$<$1.2$\times$10$^{-17}$	&$<$6.3$\times$10$^{41}$	& $<$ 2.8	&this work	\\
Q2222$-$0946 	&2.3543             	&20.50$\pm$0.15			&$-$0.46	&$-$0.99	&$-$0.51	&8.7$\pm$2.6$\times$10$^{-17}$	&38.5$\pm$1.1$\times$10$^{41}$	&17.1$\pm$5.1	&this work	\\
Q2313$-$3704 	&2.1821             	&20.48$\pm$0.10			&$<-$1.29	&$-$1.70 			&$-$1.52			&$<$1.1$\times$10$^{-17}$	&$<$4.1$\times$10$^{41}$	& $<$ 1.8	&this work	\\
Q2350$-$0052	&2.6147			&21.30$\pm$0.10			&$<-$2.10	&$-$2.23	&$-$1.97	&$<$1.5$\times$10$^{-17}$	&$<$8.6$\times$10$^{41}$	&$<$3.8	&this work	\\
Q2352$-$0028  	&1.0318			&19.81$^{+0.14}_{-0.11}$	&$<-$0.51	&$-$0.37	&$+$0.14	&4.9$\pm$2.4$\times$10$^{-17}$	&2.8$\pm$1.4$\times$10$^{41}$	&1.3$\pm$0.6	&this work	\\
\hline\hline 				       			 	 
\end{tabular}			       			 	 
\end{center}			       			 	 
\vspace{0.2cm}
\begin{minipage}{140mm}
{\bf $^a$:} References for \lognhi\ and abundance ratios are provided in Section 3 for the detected systems and in the appendix for the undetected ones. \\
{\bf $^b$:} The 2.5-$\sigma$ upper limits for non-detections are computed for an unresolved source spread over 32 spatial pixels and spectral FWHM = 6 pixels = 9 \AA. \\
\end{minipage}
\end{table*}

In the sample presented here, we report the three detections out of 16 absorbers. 
With the caveat that the sample sizes are still small, we attempt to assess which selection criteria better optimise the \ha\ emission line detections from both this sample and the earlier results from our survey for galaxy counterparts to DLAs and sub-DLAs.

When combining the data of P\'eroux et al. (2011a) and the sample presented in this paper, we note that the mean seeing for the combined cubes is 0.9" and that this seems to play little role in the likelihood of the detections: 25\%\ (two out of eight) of the detections are done in data at $>$0.9" while 21\%\ (three out of 14) are found in cubes with seeing $\leq$0.9". Natural Guide Star Adaptive Optics data, however, seem to improve slightly the detection rate from 14\%\ for no AO (2 out of 14) to 37\%\ (three out of eight), because of slight increase in sensitivity.

P\'eroux et al. (2011a) reported the detections of one DLA and one sub-DLA out of two DLAs and four sub-DLAs. Here, we have detected two DLAs and one sub-DLA out of 11 DLAs and five sub-DLAs. Combined together, this makes the search for sub-DLAs slightly more successful (33\%) than the one for DLAs (23\%), which goes in-line with findings that the former might represent more massive galaxies, with higher SFR (Khare et al. 2007, Kulkarni et al. 2007). We note however that the fraction of sub-DLAs is higher at \zabs$\sim$1 (seven out of 10) than at \zabs$\sim$2 (two out of 12), so that it is difficult to disentangle the effect of column density from a difference in redshift. 

All of the quasar absorbers presented here or in P\'eroux et al. (2011a) have [Zn/H] abundance measurements or limits derived from high-resolution spectroscopy as described in Table~\ref{t:results}. Treating limits as measures, we can compute the success rates for systems with [Zn/H] above or below the mean metallicity of the sample [Zn/H]=$-$0.65. We find 30\%\ (four out of 12) chance of detecting absorbers with [Zn/H]$>-$0.65 for 10\%\ (1 out of 10) to detect those with [Zn/H]$\leq -$0.65. Excluding limits from the sample, the mean metallicity is then [Zn/H]=$-$0.50. We find 28\%\ (two out of seven) chance of detecting absorbers with [Zn/H]$>-$0.50 for 0\%\ (zero out of three) to detect those with [Zn/H]$\leq -$0.50. In other words, using emission lines and looking within 6 arcsec of quasar sightlines, absorbers are seen four of five times when the abundances of Zn or Si are $>$1/5 solar. Four out of five detections have \ha\ SFRs above 1.3 M$_{\odot}$/yr, which is a lower limit as the data for Si are incomplete. On the contrary, four systems that meet this abundance minimum do not show detections. Three of these have detection limits near our minimum detections. One of  these is Q1323$-$0021 which has an early type galaxy within 1" with yet unconfirmed redshift (Chun et al. 2010, Rao et al. 2011). Therefore, as advocated by others before (M\o ller et al. 2002), selecting systems with higher metallicity might increase the likelihood to detect \ha\ emission, which can be easily understood in terms of higher SFR in these systems. 

Using the ratio of iron-peak elements which are mildly depleted (Zn) and refractory (Fe), one gets indications of the dust content of quasar absorbers (Pettini et al. 1994). For the undetected systems, we derive a mean $<$[Zn/Fe]$>$=$+$0.66$\pm$0.96 (treating limits as measures) for 16 systems while the five detected systems have $<$[Zn/Fe]$>$=$+$0.42$\pm$0.41. Hence, we see a hint of a trend whereby less dusty systems are more likely to be detected. Incidentally, we note that the mean alpha-peak to iron-peak elements ratio are $<$[Si/Fe]$>$=$+$0.57$\pm$0.82 for the nine undetected systems which have [Si/H] measurements, and $<$[Si/Fe]$>$=$+$0.45$\pm$0.41 for the four corresponding detected ones. We note that some of the individual [Si/Fe] values are high i.e. [Si/Fe]=$+$1.20 in the absorber towards Q2102$-$3553.

However, looking at the detection rates, we find that the factor which is the most discriminatory is the actual redshift (and therefore SINFONI observing band) of the absorber. At face value, we note that 40\%\ (four out of ten; that is, two detections of five in P\'eroux et al. (2011a) and two of five in this work) of the \zabs$\sim$1 systems are detected while only 8\%\ (one out of 12;  that is zero out of two in P\'eroux et al. (2011a) and one of 11 in this work) of the \zabs$\sim$2 are found. 
These results are in line with recent reports from Bouch\'e et al. (2011) indicating that the success rate of detections of \mgii-selected absorbers at z$\sim$2 is much lower than the detection rate at z$\sim$1 (Bouch\'e et al. 2007), taking into account the increase in luminosity distance. Bouch\'e et al. (2011) advocate that these results cannot be explained by the galaxies hiding within the quasar image or being outside the SINFONI field of view. In the case of \hi-selected absorbers, we note, indeed, that the impact parameters probed by our observational approach can be quite low (down to 6kpc in the case of Q2222$-$0946) while the search radius is large compared with expectations from DLA/sub-DLA galaxies (up to 39kpc in the case of Q1009$-$0026, see P\'eroux et al. 2011b; although see Chen et al. 2010 too). Still, in the case of Q0001$-$0159, Van der Werf, Moorwood \& Bremer (2000) have reported an unconfirmed galaxy candidate which falls just outside our SINFONI field. Nevertheless, the search radius at \zabs$\sim$1 and \zabs$\sim$2 are comparable (41 and 43 kpc, respectively) and cannot explain alone the difference in detection rates. These differences with redshift led Bouch\'e et al. (2011) to conclude that z$\sim$2 \mgii\ absorbers reside in halos smaller than z$\sim$1 systems based on redshift evolution arguments. It is important to note that it can well be that some of the targeted galaxies are bright in the continuum but not strong emitters or are dwarf galaxies with low metallicities and low star formation rate which at z$\sim$2 have not yet merged with bigger galaxies.

\subsection{Star Formation Rate}

\begin{table*}
\begin{center}
\caption{Properties of the two previously detected and three newly detected DLAs and sub-DLAs in the SINFONI cubes.}
\label{t:detections}
\begin{tabular}{cccccccccc}
\hline\hline
Quasar 		  &$\Delta v^a$  &$\Delta$ RA  &$\Delta$ Dec &projected dist &phys dist  	  &$\Sigma_{SFR}$ &\nii/\ha & 12$+$log(O/H) &Reference\\
			  &[km/s]		  	     &["]		   &["]			  & ["] 	&[kpc]	   	&[M$_{\odot}$/yr/kpc$^2$] & & &		 	 \\
\hline
Q0302$-$223				&1			&$+$0.50   	&$+$3.12  	&3.16	&25		&0.13	&$<$0.34	&$<$8.6	&P\'eroux et al. 2011a\\
Q0452$-$1640 			&10			&$-$0.125	&$+$2.00		&2.00	&16		&0.49	&0.12$^{+0.10}_{-0.05}$	&8.4$\pm$0.1		&this work\\
Q1009$-$0026   			&125		&$-$5.01  		&$+$0.37 		 &5.01	&39		&0.31	&0.48$^{+0.56}_{-0.27}$	&8.7$\pm$0.2		&P\'eroux et al. 2011a\\
Q2222$-$0946 			&30            	&$+$0.50		&$+$0.50		&0.70	&6		&1.97	&$<$0.05&$<$8.2	&this work\\
Q2352$-$0028 			&10			&$+$1.50 		&$+$0.00		&1.50	&12		&0.24	&0.12$^{+0.24}_{-0.08}$	&8.4$\pm$0.3		&this work\\
\hline\hline 				       			 	 
\end{tabular}			       			 	 
\end{center}			       			 	 
\begin{minipage}{140mm}
{\bf $^a$:} Observed velocity shift corresponding to $\Delta z$=\zabs$-$\zem\ of the detected galaxies. \\
\end{minipage}
\end{table*}

\begin{figure*}
\begin{center}
\includegraphics[height=8cm, width=8cm, angle=-90]{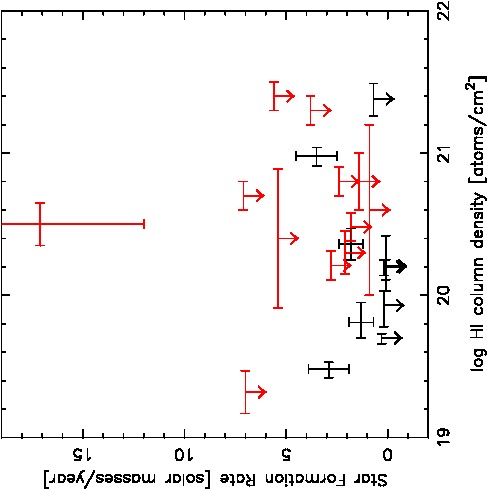}
\includegraphics[height=8cm, width=8cm, angle=-90]{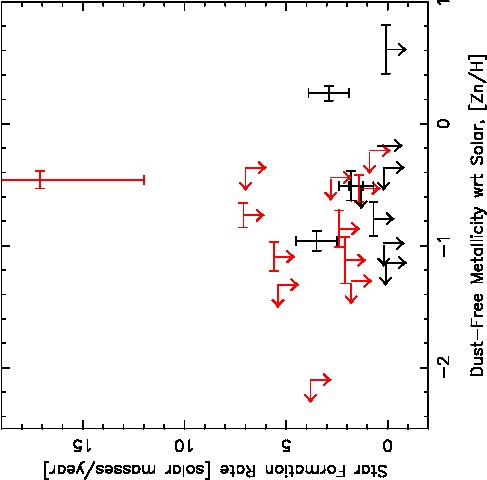}
\caption{{\bf Left Panel:} Star formation rate as a function of \hi\ column density for the full sample. The black points are for systems at z$\sim$1 and the red ones for systems at z$\sim$2. There is no clear correlation between these quantities in our sample. {\bf Right Panel:} Star formation rate as a function of metallicity with respect to solar as traced by dust-free [Zn/H]. We note that the detections were made among the most metal-rich systems. }
\label{f:sample}
\end{center}
\end{figure*}

In this section, we provide more details on the three \ha\ detections presented in this work. We estimate the \ha\ fluxes by extracting emission lines within a given radius. For the absorber at \zabs=1.0072 towards Q0452$-$1640, we use a radius of 13 pixels or 3.25" (seeing=1.1"); for the \zabs=2.3543 system toward Q2222$-$0946 and the \zabs=1.0318 system toward Q2352$-$0028, we used 5 pixels or 1.25" to avoid contamination by the quasar nearby (seeing=0.6" and 0.7", respectively). The fluxes are estimated from a Gaussian fit to the emission lines and resulting values are listed in Table~\ref{t:results}. Using the H-$\alpha$ luminosity, we then derive the star formation rate (SFR) assuming the Kennicutt (1998) flux conversion corrected to a Chabrier (2003) IMF. These estimates are not corrected for dust extinction. Some of the SFR limits reached in that way are among the most sensitive limits obtained to date.

For the DLA with \lognhi=20.98$^{+0.06}_{-0.07}$ at \zabs=1.0072 towards Q0452$-$1640, we derive F(\ha)=14.5$\pm$4.3$\times$10$^{-17}$ erg/s/cm$^2$. This corresponds to L(\ha)=8.0$\pm$2.4$\times$10$^{41}$ erg/s which leads to a SFR estimate of SFR=3.5$\pm$1.0 M$_{\odot}$/yr. For the DLA  with \lognhi=20.50$\pm$0.15 at \zabs=2.3543 towards Q2222$-$0946, we derive F(\ha)=8.7$\pm$2.6$\times$10$^{-17}$ erg/s/cm$^2$. This corresponds to L(\ha)=38.5$\pm$1.1$\times$10$^{41}$ erg/s which leads to a SFR estimate of SFR=17.1$\pm$5.1 M$_{\odot}$/yr. We therefore report a higher SFR than Fynbo et al. (2010) who detected this object by triangulation of X-Shooter spectra in the field, but we note that their data do not cover the actual galaxy leading to important "slit loss" and that their calibration was artificially decreased by 30\% to match the quasar flux in the Sloan spectrum. Therefore, their SFR estimate of $>$10 M$_{\odot}$/yr should be viewed as a lower limit and is consistent with our result. This object is the only z$\sim$2 detected object in our \hi-selected absorbers SINFONI search and one should note that the SFR is considerably higher than the one from other detections at z$\sim$1, probably because of the SFR-M$_*$ sequence. Finally, for the sub-DLA with \lognhi=19.81$^{+0.14}_{-0.11}$ at \zabs=1.0318 towards Q2352$-$0028, we derive F(\ha)=4.9$\pm$2.4$\times$10$^{-17}$ erg/s/cm$^2$. This corresponds to L(\ha)=2.8$\pm$1.4$\times$10$^{41}$ erg/s which leads to a SFR estimate of SFR=1.3$\pm$0.6 M$_{\odot}$/yr. 

For all the remaining absorbers searched for, we reach stringent star formation limits. The 2.5-$\sigma$ upper limits for non-detections are computed for an unresolved source spread over 32 spatial pixels and spectral FWHM = 6 pixels = 9 \AA. These limits are summarised in Table~\ref{t:results}. We note that the resulting luminosity limits vary from one field to another, complicating the interpretation of detection rates.

In Figure~\ref{f:sample}, we present the derived SFR as a function of both the \nhi\ column density and metallicity for the detections and the non-detections presented in this work. The black points are for systems at z$\sim$1 and the red ones for systems at z$\sim$2. We note that some of the limits at z$\sim$2 would not be stringent enough to reach some of the detections at z$\sim$1. Within the small number of objects currently observed, we see no clear correlation between the quantity plotted, although we note that the detections were made among the most metal-rich systems. It is possible that the absence of trends is due to a dust bias where at high metallicity only systems of low \nhi\ column density are observed, and at high \nhi, only objects of low metallicity are observed (see e.g. Vladilo \& P\'eroux 2005). We note that the detection of the galaxy towards Q2222$-$0946 (SFR=17.1$\pm$5.1 M$_\odot$ /yr) is well above the typical detection limits.

 \subsection{Star Formation Rate per Unit Area}

\begin{figure}
\begin{center}
\includegraphics[height=7cm, width=10cm, angle=0]{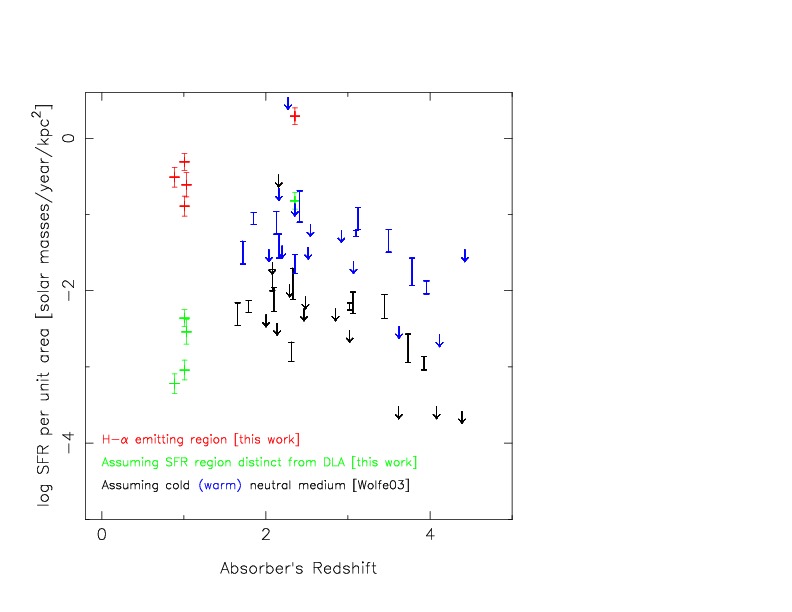}
\caption{Star formation rate per unit area in quasar absorbers as a function of redshift. The black and blue points are inferred from the two cooling rates measured by the strength of \cii$^*$ \l1335.7 absorption (Wolfe, Prochaska \& Gawiser 2003). Predictions vary according to whether a cold neutral medium (CNM, in black) or a 
warm neutral medium (WNM, in blue) is assumed. These results can be directly compared with our measurements of $\Sigma_{SFR}$ in the 5 detections presented here in red for the \ha\ emitting regions and in green when we assume absorbers are passive probes of the radiation field at the impact parameter distance from the detected central core of star formation traced by \ha.}
\label{f:z_Wolfe}
\end{center}
\end{figure}

\begin{figure*}
\begin{center}
\includegraphics[height=7.cm, width=8.5cm, angle=0]{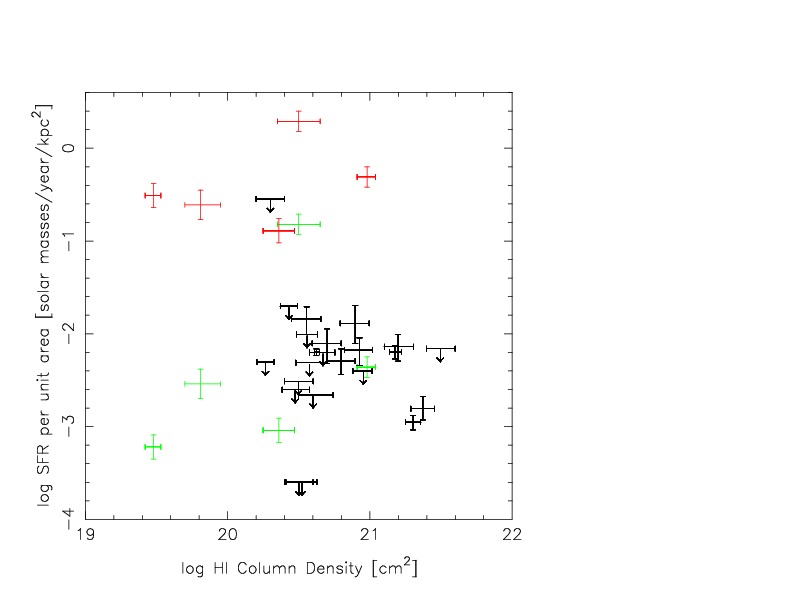}
\includegraphics[height=7.cm, width=8.5cm, angle=0]{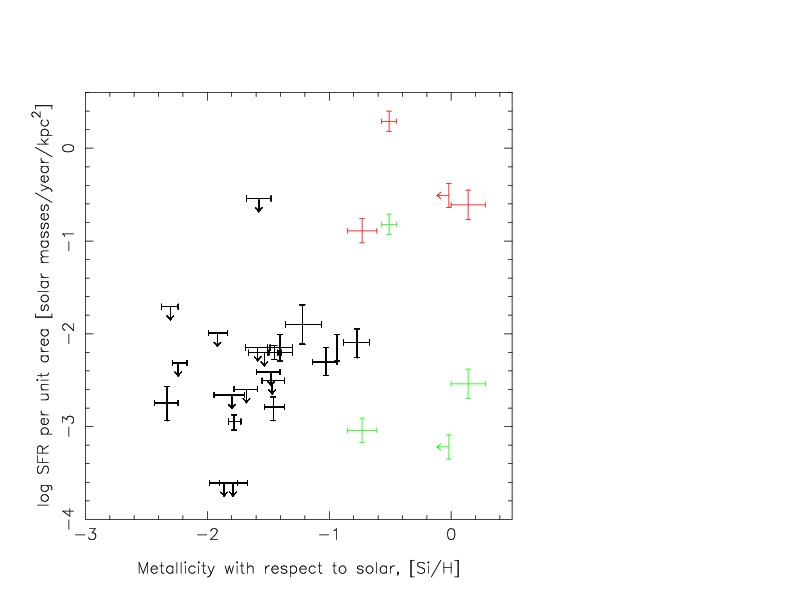}
\caption{{\bf Left panel:} Star formation rate per unit area in quasar absorbers as a function of \hi\ column density. {\bf Right panel:} Star formation rate per unit area in quasar absorbers as a function of metallicity with respect to solar as traced by [Si/H]. The black points are inferred from \cii$^*$ measurements assuming a cold neutral medium phase which is favoured by Wolfe, Gawiser \& Prochaska (2003). Our measures of the same quantities from the SINFONI observations are presented in red for the \ha\ emitting regions and in green when we assume absorbers are passive probes of the radiation field at the impact parameter distance from the detected central core of star formation traced by \ha. }
\label{f:NHI_Wolfe}
\end{center}
\end{figure*}

Recently, Wolfe, Prochaska \& Gawiser (2003) have proposed a technique that measures star formation rates in DLAs inferring the heating rate by equating
it to the cooling rate measured by the strength of \cii$^*$ \l1335.7 absorption. In this way, they compute the SFR per unit area for a sample of almost 30 DLAs in which the dust-to-gas ratio has been inferred from element depletion patterns in two different models: one in which the line of sight passes through cold neutral medium (CNM, T $\sim$ 80 K) and
warm neutral medium (WNM, T $\sim$ 8000 K) - the so-called CNM model - and another one where the line of sight passes only through WNM gas - the WNM model. They obtain SFR per unit area equals 10$^{-2.2}$ M$_{\odot}$ yr$^{-1}$ kpc$^{-2}$ (similar to the one measured in the Milky Way interstellar medium) for the
CNM solution and 10$^{-1.3}$ M$_{\odot}$ yr$^{-1}$ kpc$^{-2}$ for the WNM solution. Since the heating rate is
proportional to the product of the dust-to-gas ratio, the grain photoelectric heating efficiency, and the
SFR per unit area, the WNM solution lead to higher estimate of the SFR per unit area. One of the systems in that sample (Q1234$+$0758) is part of the current SINFONI survey but it remains undetected in \ha. Nevertheless, for the five detections of our survey, both the SFR and size of \ha\ emitting regions are measured providing a robust estimate of the star formation rate per unit area in the galaxies. These values are tabulated in Table~\ref{t:detections} and shown in red in Figure~\ref{f:z_Wolfe} as a function of absorber's redshift. These show values which are more in line with results from the WNM model but one should bear in mind that our detection limits might be above values expected from CNM model. Alternatively, we can also compute the SFR per unit area assuming a model in which the absorbers are passive probes of the radiation field at some distance taken to be the impact parameter from the central core of star formation detected in \ha\ (Rafelski, Wolfe \& Chen 2011). These results are shown in green in Figure~\ref{f:z_Wolfe} and can be viewed as lower limits to a comparison with results from \cii$^*$ estimates. They show values which are in agreement with results from the CNM model.

In addition, in Figure~\ref{f:NHI_Wolfe}, we present these measures as a function of \nhi\ column density and metallicity traced by [Si/H] and compare it with the sample of Wolfe, Gawiser \& Prochaska (2003) assuming CNM as favoured by the authors. Our data probe lower column densities and because of our selection criterion, higher metallicities. 

\subsection{Velocity Shift and Impact Parameter}

\begin{figure*}
\begin{center}
\includegraphics[height=5cm, width=5cm, angle=-90]{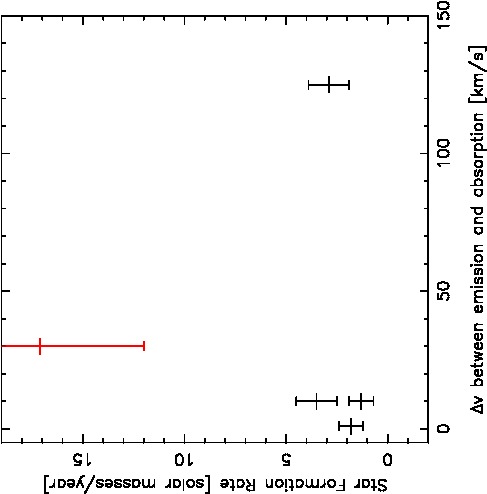}
\includegraphics[height=5cm, width=5cm, angle=-90]{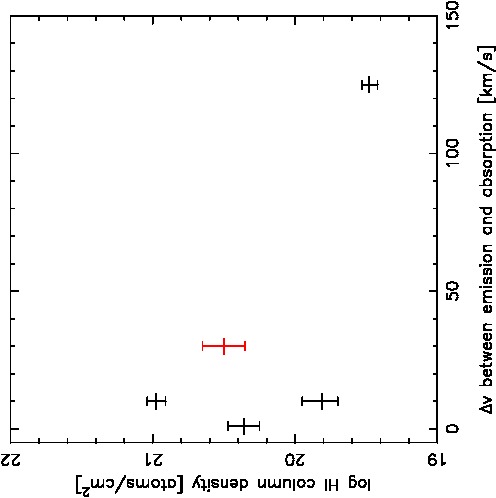}
\includegraphics[height=5cm, width=5cm, angle=-90]{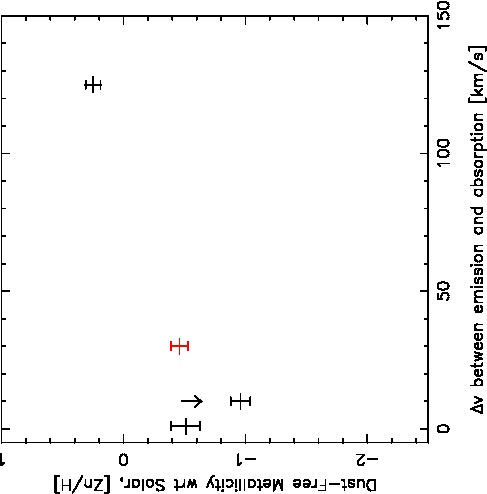}
\includegraphics[height=5cm, width=5cm, angle=-90]{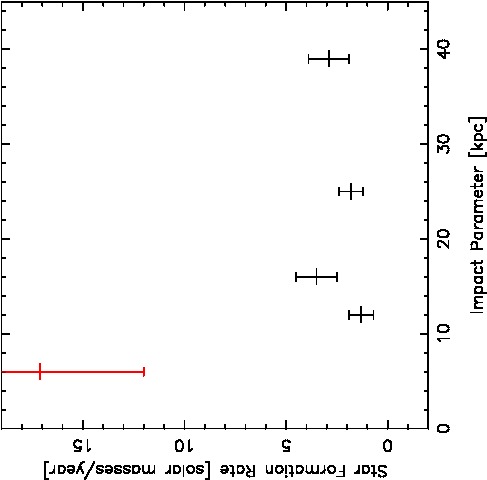}
\includegraphics[height=5cm, width=5cm, angle=-90]{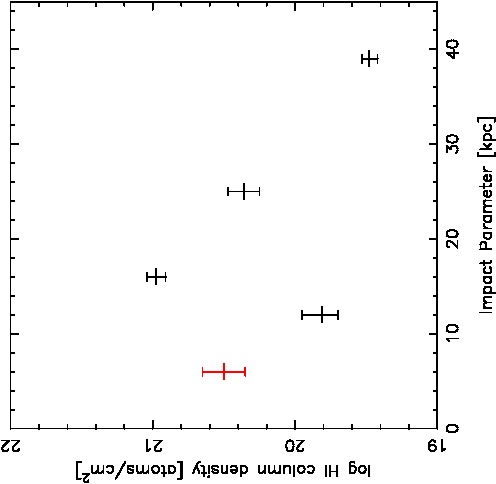}
\includegraphics[height=5cm, width=5cm, angle=-90]{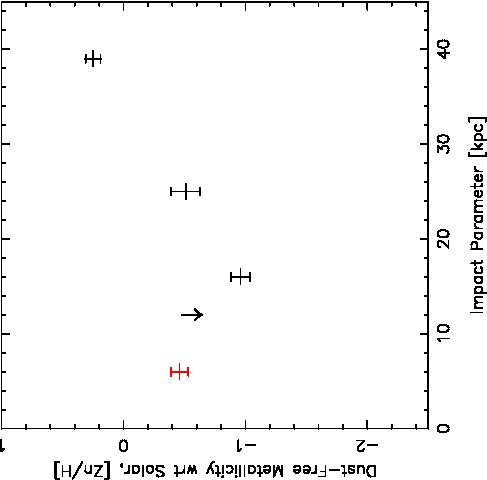}
\caption{{\bf Top Panels:} Star formation rate, log of \hi\ column density and metallicity with respect to solar as traced by dust-free [Zn/H] as a function of the velocity shift between the emitting galaxy and the absorbing material seen along the quasar's line-of-sight for the 5 systems detected in our SINFONI survey. The black points are for systems at z$\sim$1 and the red one for the system at z$\sim$2. {\bf Bottom Panels:} Same quantities as a function of impact parameter. }
\label{f:detected}
\end{center}
\end{figure*}

Thanks to the \ha\ maps, we are able to accurately report the sky position and impact parameters between the quasar's line-of-sight and the center of the detected galaxies. Table~\ref{t:detections} summarises these properties. Once again, one can see that IFU observational approach allows to probe impact parameters which can be quite small (down to 6kpc in the case of the absorber towards Q2222$-$0946). The three detections reported here have indeed impact parameters lower than the one found in P\'eroux et al. (2011b). In addition, we compute the observed velocity shift between the absorber and the detected galaxies, $\Delta v$ with an accuracy of about 30 km/s. In some cases, the large values observed suggest that some of the absorption features observed might be due to material outflowing from the galaxy itself. If true, the kinematic observations would allow to make an estimate of the rate of the outflowing material. In the case of Q2222$-$0946, Fynbo et al. (2010), who rely on a X-Shooter spectrum for both the emission and absorption redshift determination, derive from the \ha, an emission redshift z=2.3537, resulting in $\Delta v$=35$\pm$10 km/s (Fynbo, private communication). Our estimate is in good agreement with this value. It is interesting to note however, that their emission redshift determination based on \oiii\ z=2.35406 is in better agreement with the redshift of low-ionisation absorption lines.

In Figure~\ref{f:detected}, we plot SFR, log \nhi\ \& metallicity, as a function of both the velocity shift, $\Delta v$ , between the galaxy and the absorbing material seen along the quasar's line-of-sight and the impact parameter. In spite of the very small size of the sample,
the range of impact parameter is relatively large, spanning from few kpc, to more than 125 kpc. So, in principle, we may probe the outskirts 
and surroundings of the main galaxy. The effect the most expected in this Figure (Rao et al. 2011) would be a decrease of \nhi\ with impact parameter and there is indeed a hint of that effect (Spearman's rank correlation coefficient r$_s$=$-$0.56), to be confirmed by a larger data set. Similarly, we see no clear hint of decreasing SFR with impact parameter, although the highest SFR is at the smallest impact parameter. The metallicity [Zn/H] is more strongly correlated with $\Delta v$  (r$_s$=0.81) and the impact parameter (r$_s$=0.70). Assuming some of the material is outflowing, $\Delta v$, in combination with the known orientation of galaxy from SINFONI observation, will allow to interpret the high-resolution absorption profile kinematics for these objects.

\subsection{Metallicity}

\begin{figure}
\begin{center}
\includegraphics[height=8cm, width=10cm, angle=0]{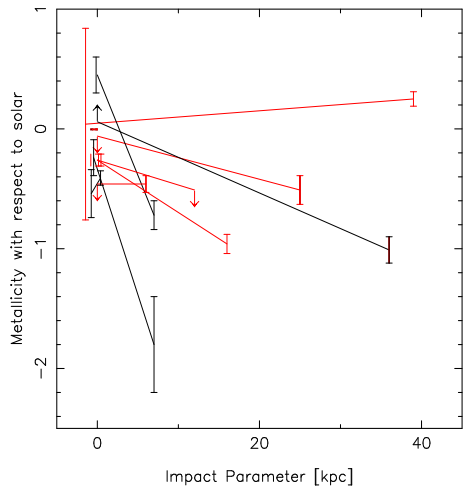}
\caption{Metallicity with respect to solar measured in emission at zero impact parameters and in absorption at given impact parameters in kpc. The two metallicities arising from the same galaxy are linked by a line. The data from our survey (shown in red) have more than doubled the number of systems for which such measures are possible. The arrows indicate upper and lower limits on the measure of the metallicities. The errors on the emission metallicities are artificially offset from zero impact parameter for clarity.}
\label{f:Gradients}
\end{center}
\end{figure}

\begin{table*}
\begin{center}
\caption{Metallicity with respect to solar measured in absorption at given impact parameter and in emission. }
\label{t:Gradients}
\begin{tabular}{ccccccc}
\hline\hline
Quasar 		  &phys dist  	&Absorption Abundance  &Ion&Emission Metallicity$^a$ &Gradients &Reference\\
			  &[kpc]	   	&[X/H] &X &12$+$log(O/H) &[dex/kpc]&		 	 \\
\hline
HS1543$+$5921	&0.4 &$-$0.41$\pm$0.06          &S  &$-$0.54$\pm$0.20      &$+$0.32$\pm$0.21 &Bowen et al. 2005\\
Q1009$-$0026		&39  &$+$0.25$\pm$0.06         &Zn  &$+$0.04$\pm$0.80   &$+$0.01$\pm$0.80 &This work\\
AO0235$+$164	&7     &$-$1.80$\pm$0.40          &Fe &$-$0.24$\pm$0.15	  &$-$0.22$\pm$0.43 &Chen et al. 2005\\
Q0302$-$223		&25  &$-$0.51$\pm$0.12           &Zn &$<-$0.06		            &$>-$0.02 &This work\\
PKS0439$-$433	&7     &$-$0.72$\pm$0.12           &Fe &$+$0.45$\pm$0.15   &$-$0.17$\pm$0.19 &Chen et al. 2005\\
Q0827$+$243		&36   &$-$1.01$\pm$0.11           &Fe &$>+$0.06    		   &$<-$0.03 &Chen et al. 2005\\
Q0452$-$1640		&16   &$-$0.96$\pm$0.08           &Zn &$-$0.26$\pm$0.01    &$-$0.04$\pm$0.08 &This work\\        
Q2222$-$0946		&6     &$-$0.46$\pm$0.07            &Zn&$<-$0.46		   &$>-$0.00 &This work\\
Q2352$-$0028		&12   &$<-$0.51 		            &Zn&$-$0.26$\pm$0.03    &$<-$0.02 &This work\\
\hline\hline 				       			 	 
\end{tabular}			       			 	 
\end{center}			       			 	 
\begin{minipage}{140mm}
{\bf $^a$:} The emission metallicities are derived from R$_{23}$ (Pagel et al. 1979) except for objects studied in this work where we used N2 (Pettini \& Pagel 2004). \\
\end{minipage}
\end{table*}

Using the N2-parameter (Pettini \& Pagel 2004) based on \nii $\lambda$ 6585/H-$\alpha$ ratio, we can derive an estimate of the emission metallicity. Such observations of the absorption and emission metallicities in the same high-redshift object allow to trace the two different gas phases (neutral versus ionised gas). 
For the DLA towards Q0452$-$1640, we find 12+log(O/H)=8.4$\pm$0.1, i.e. slightly less than solar\footnote{the solar abundance is 12+log(O/H)$_{\odot}$=8.66 (Asplund et al. 2004).} compared to the one-tenth solar absorption metallicity reported in absorption ([Zn/H]=$-$0.96$\pm$0.08). For the DLA towards Q2222$-$0946, we find 12+log(O/H)$<$8.2 from a non-detection, i.e. below solar compared with the one-third absorption metallicity ([Zn/H]=$-$0.46$\pm$0.07). Finally, for the sub-DLA towards Q2352$-$0028, 12+log(O/H)=8.4$\pm$0.3, i.e. slightly less than solar compared to the less than one-third solar absorption metallicity reported in absorption ([Zn/H]$<$$-$0.51). These values are summarised in Table~\ref{t:detections}.  

Following our investigations in Paper I, we compare the metallicity with respect to solar as a function of impact parameters. Table~\ref{t:Gradients} lists the absorption and emission abundances for the quasar absorbers for which both measures are available. Figure~\ref{f:Gradients} illustrates the observed gradients derived from these observations.The two metallicities arising from the same galaxy are linked by a line. The data from our survey have more than doubled the number of systems for which such measures are possible. Out of the eight galaxy counterparts to quasar absorber, two have abundances at the impact parameter which are slightly higher than in the center of the galaxy, although the statistical significance of the results is limited given the large error estimates in the emission metallicity measurements (Table~\ref{t:Gradients}). These results could be due to a patchy metals distribution across the galaxies, but we note that "inverted gradients" have been recently reported in the literature as a signature of primordial gas infall onto galaxies (Cresci et al. 2010). Based on a larger sample of metallicity gradients from the MASSIV survey (Contini et al. 2012), Queyrel et al. (2012) suggest that the gas infall might be due to either interactions or cold gas accretion. We also note that using lensed galaxies to study metallicity gradients allows to probe smaller scales (Jones et al. 2010).

\section{Conclusion}

In this study, we report three additional SINFONI detections of DLA/sub-DLAs, two of which are new identifications. Together with our earlier results of P\'eroux et al. (2011b), we therefore have detected five absorbers with \lognhi$>$19.0, in comparison with 20 such detections at all redshifts reported over several years of dedicated searches (Rao et al. 2003, Chen et al. 2003, Rosenberg et al. 2003, Fynbo et al. 2010, Fynbo et al. 2011). With the limitation of the small number statistics, we find that selecting systems with higher metallicity might increase the likelihood to detect \ha\ emission. In line with the results of a similar study of \mgii\ absorbers obtained by Bouch\'e et al. (2011), we find that the detection rate is significantly lower at \zabs$\sim$2 than at \zabs$\sim$1.

For the five absorbers which are detected, we are able to measure SFR, SFR per unit area, impact parameter to the quasar line-of-sight, velocity shift between absorption and emission redshift and emission metallicities. We derive star formation rates (SFR) of a few M$_{\odot}$/yr at \zabs$\sim$1 and SFR=17 M$_{\odot}$/yr for the DLA at \zabs$\sim$2. Our estimates of the SFR per unit area are compared with results from Wolfe, Prochaska \& Gawiser (2003) who inferred these from CII$^*$ absorption. Our observations show values which are more in line with results from their warm neutral medium model than the cold one. The detected galaxies are at impact parameters ranging from 6 to 12 kpc from the quasar's line-of-sight. We note again that the IFU technique allows detections at small impact parameter to the quasar. Indeed, one of the detection reported here is at \zabs$\sim$2 and was detectable in spite of being only 0.7" away from the quasar. As found by others (Rao et al. 2011), we see a decrease of \nhi\ with impact parameter. Finally, our observations of the absorption and emission metallicities in the same high-redshift object allow to trace the two different gas phase (neutral versus ionised gas). The data from our survey have more than doubled the number of systems for which such measures are possible. Out of the eight galaxy counterparts to quasar absorbers, two might have "inverted gradients", were the abundance at the impact parameter is higher than in the center of the galaxy.

All of these absorbers have high-resolution quasar spectra and in future papers, we will be able to undertake detailed studies of the absorption/emission metallicities (i.e. Kacprzak et al. 2011) and kinematics of these objects. The growing amount of data which is currently being assembled will allow to undertake direct comparison on the nature of DLAs and sub-DLAs with expectations from dedicated cosmological models (Tescari et al. 2009, Hong et al. 2010, Cen 2010).
 
\section*{Acknowledgements}
We would like to thank the Paranal and Garching staff at ESO for performing the observations in Service Mode and the instrument
team for making a reliable instrument. CP thanks Johan Fynbo, Max Pettini, Thierry Contini, Chris Howk and Art Wolfe for useful discussions and Thomas Ott for developing and distributing the QFitsView software. VPK acknowledges partial support from the U.S. National Science Foundation grants AST-0908890 (PI: Kulkarni). This work has benefited from support of the "Agence Nationale de la Recherche" with reference ANR-08-BLAN-0316-01.

\appendix

\section{Note on Individual Fields with No Detections}

\subsection{Q0001$-$0159, \zabs=2.0950 (0.871, 2.1539)}

Smith, Cohen \& Bradley (1986) first reported two absorbers along this line-of-sight at \zabs=2.0950 and 2.1539. 

For the first system, Prochaska \& Wolfe (1999) derive from HIRES spectroscopy: \lognhi=20.70$\pm$0.10, [Al/H]=$-$1.518$\pm$0.106, [Si/H]=$-$0.842$\pm$0.102, [Cr/H]=$-$1.632$\pm$0.135, [Ni/H]=$-$1.808$\pm$0.114, [Fe/H]=$-$1.703$\pm$0.103 and [Zn/H]=$-$0.755$\pm$0.104.

For the other system, Prochaska \& Wolfe (1999) derive from HIRES spectroscopy: \lognhi=20.30$\pm$0.10, [Al/H]=$-$1.615$\pm$0.101, [Si/H]=$-$1.534$\pm$0.101, [Cr/H]$<-$1.034, [Ni/H]$<-$1.601, [Fe/H]=$-$1.915$\pm$0.105 and [Zn/H]=$<-$1.049.

Meyer \& Roth (1990) also identify a \mgii\ and \feii\ system at \zabs=0.871.

Bunker et al. (1999) used long-slit K-band spectroscopic on UKIRT to try and detect the galaxies responsible for these two absorbers. They did not find any, leading to upper limits of F(\ha)$<$10.5$\times$10$^{-17}$ erg/s/cm$^2$, i.e. SFR$<$7.23 h$^{-2}$M$_{\odot}$/yr at \zabs=2.0950 and F(\ha)$<$10.0$\times$10$^{-17}$ erg/s/cm$^2$, i.e. SFR$<$7.33 h$^{-2}$M$_{\odot}$/yr at \zabs=2.1539. 

However, Van der Werf, Moorwood \& Bremer (2000) report a galaxy candidate consistent with $z$=2.1539 from IR narrow-band flux at \ha\ with K broad-band continuum measurement. The object, detected 7.7" south-west of the quasar has F(\ha)=1.8 $\times$10$^{-16}$ erg/s/cm$^2$ corresponding to SFR=23 h$^{-2}$ M$_{\odot}$/yr. Unfortunately, this region is just outside our SINFONI field. Colbert \& Malkan (2002) also used NICMOS to image this field and reach H magnitudes of 20.9 at \zabs=2.0950 and 21.0 at \zabs=2.1539 without detecting any galaxy-candidates.

Therefore, we do not detect either the galaxy responsible for the system at \zabs=2.1539 nor the main target at \zabs=2.0950 in our SINFONI data. For the latter, we derive F(\ha)$<$4.5$\times$10$^{-17}$ erg/s/cm$^2$, i.e. SFR$<$11 M$_{\odot}$/yr. Finally, the \ha\ emission line of the \mgii\ system at \zabs=0.871 is not covered by our K-band data.

\subsection{Q1211$+$0902, \zabs=2.5841 (1.2650, 2.33, 3.1919)}

This high-redshift quasar was first reported by Hazard, McMahon \& Morton (1987) but is also part of the Data Release 3 of the Sloan Digital Sky Survey (Richards et al. 2006). From an early AAT (Anglo-Australian Telescope) spectrum, the former authors suspected the presence of a strong \nhi\ absorber at \zabs=2.5841. The \nhi\ column density was measured to be \lognhi=21.40$\pm$0.10 by Ledoux et al. (2006). The \znii\ metallicity of that system was first studied by Pettini et al. (1997), the \civ\ profile associated was studied in detail by Fox et al. (2007), while Prochaska et al. (2007) used ESI/HIRES Keck Telescope spectra to measure the abundances of several lines in this system. They derive: 
[Al/H]$>-$1.82, [Si/H]=$-$1.05$\pm$0.11, [Cr/H]=$-$1.48$\pm$0.15, [Fe/H]=$-$1.65$\pm$0.13, [N/H]=$-$2.22$\pm$0.12 and [Zn/H]=$-$1.09$\pm$0.12. As usual in quasar absorbers, the [N/H] values are low (Pettini et al. 2002). The strong lines are spread over $\sim$600 km/s. The \mgii\ lines are not covered by these spectra. Petitjean et al. (2006) have reported the upper limit of the molecular hydrogen content of that absorber to be log f$<-$5.69. The abundance pattern in this system is interesting, since both Si and Fe are more depleted with respect to Zn, suggesting the presence of dust, yet no H$_2$ was detected.

Lanzetta et al. (1991) reported from a Las Campanas 2.5m-telescope spectrum, an additional absorber at \zabs=2.33 which was believed to be below the canonical DLA definition. Natarayan et al. (2008) have also found an EW(\mgii\ 2796)=0.083$\pm$0.07\AA\ \mgii\ system at \zabs=1.2650. Finally, Ryabinkov, Kaminer \& Varshalovich (2003) mention an additional \civ\ and \siii\ absorption feature at \zabs=3.1919.

Lowenthal et al. (1995) have taken Fabry-Perot observations of that field but do not detect the galaxy counterpart to the main absorber at \zabs=2.5841, providing an upper limit F(\lya)$<$3 $\times$10$^{17}$ ergs/s/cm$^2$ at 3-$\sigma$, corresponding to SFR$<$1 M$_\odot$/yr. Colbert \& Malkan subsequently used NICMOS imaging in this field but do not report detection of the galaxy responsible for the quasar absorber. Similarly, we do not detect the \ha\ line in our SINFONI data. We derive F(\ha)$<$2.3 $\times$10$^{17}$ ergs/s/cm$^2$, corresponding to SFR$<$5.6 M$_\odot$/yr.

For the non-prime targets, only the \zabs=2.33 system is covered by our SINFONI data but it remains undetected. 

\subsection{Q1220$-$0040, \zabs=0.9746}

This quasar is part of the Early Release of the SDSS (Schneider et al. 2002). An absorber at \zabs=0.9746 was reported by Rao, Turnshek \& Nestor (2006). They measure \lognhi=20.20$^{+0.05}_{-0.09}$ with EW(\mgii\ 2803)=1.793$\pm$0.125\AA. Meiring et al. (2008) subsequently estimate the abundances in this absorber using the MIKE spectrograph. They use 9 components over $\Delta$ v $\sim$ 300 km/s to derive: [Zn/H]$<-$1.14 and [Fe/H]=$-$1.33$\pm$0.07. 

The galaxy responsible for this absorber is not detected in our SINFONI data.

\subsection{Q1225$+$0035, \zabs=0.7731}

This quasar is part of the Early Release of the SDSS (Schneider et al. 2002). An absorber at \zabs=0.7731 was reported by Rao, Turnshek \& Nestor (2006). They measure \lognhi=21.38$^{+0.11}_{-0.12}$ with EW(\mgii\ 2803)=1.447$\pm$0.150\AA. Meiring et al. (2006) subsequently estimate the abundances in this absorber using the Multiple Mirror Telescope (MMT). They use one component-fit to derive: [Fe/H]=$-$1.19$\pm$0.015 and [Cr/H]=$-$1.08$\pm$0.22. In addition, Nestor et al. (2008) derive [Zn/H]=$-$0.78$\pm$0.14.

Recently, based on J, H and K imaging, Rao et al. (2011) report a galaxy candidate in that field at impact parameter of 8.2" (60.8 kpc). The identification is based on proximity to the quasar, but no photometric redshift could be established. This object falls outside our SINFONI field-of-view and the galaxy responsible for this absorber is not detected in our data.

\subsection{Q1226$+$1736, \zabs=2.5576 (2.4658)}

This quasar was first discovered as part of the Large Bright Quasar Survey (LBQS). Pettini et al. (1994) used a WHT spectrum to measure \lognhi=21.52$\pm$0.10 at \zabs=2.4658 and to study the \znii\ metallicity of this absorber. 

Petitjean et al. (2000) report a H$_2$ fraction log f$<-$6.92. Prochaska et al. (2001) have reported the metallicity of that absorber based on Keck/HIRES observations, they derive: [C/H]$>-$2.89, [O/H]$>-$2.893, [Si/H]=$-$1.59$\pm$0.10, [Cr/H]=$-$1.68$\pm$0.10, [Fe/H]=$-$1.84$\pm$0.10,  [Ni/H]=$-$1.80$\pm$0.10 and [Zn/H]=$-$1.62, suggesting little dust in this system, but 1/60$^{th}$ solar abundance attributable to nucleosynthetic element production. The profiles for this system are spread over 100 km/s (Prochaska \& Wolfe 2002).

Dessauges-Zavadsky et al. (2003) have also reported a sub-DLA towards this quasar at \zabs=2.5576 with \lognhi=19.32$\pm$0.15. They fitted the absorber with 11 components over $\sim$ 200 km/s and derived: [O/H]$<-$0.77, [Si/H]=$-$0.50$\pm$0.15, [Al/H]=$-$0.61$\pm$0.15, [Fe/H]=$-$0.84$\pm$0.15, [Zn/H]$<-$0.36 and [Cr/H]$<-$1.05, indicating modest nucleosynthesis products with no indication of dust.
The low [O/Si] ratio might indicate that the system is ionised, not neutral, in O and H or that the system is Si-rich.

Warren et al. (2001) did not find candidates in that field in their NICMOS search down to H$_{AB}$=25.0. Similarly, we do not detect the \ha\ lines of any of these absorbers in the shallow SINFONI data available to us.

\subsection{Q1234$+$0758, \zabs=2.3376}

This quasar, was discovered by Foltz et al. (1987) and subsequently observed for its absorber at \zabs=2.3376 by several authors. Notably, the HD molecule was reported in that system by Varshalovich et al. (2001) to have N(HD)=1-4$\times$10$^{14}$ and the H$_2$ molecule was found by Ge, Bechtold \& Kulkarni (2001) to have N(H$_2$)=3$\times$10$^{17-19}$ depending on the precise value assumed for the Doppler parameter.  Ge, Bechtold \& Kulkarni (2001) also derived [Zn/H]=$-$0.86$\pm$0.15. This system has been later reviewed by Srianand et al. (2005) who derive: log N(H$_2$)=19.57$\pm$0.12.

The system has then been studied at higher resolution with UVES by Centuri\' on et al. (2003) who fitted 6 components over $\Delta$ v $\sim$ 140 km/s: \lognhi=20.80$\pm$0.10, [N/H]=$-$2.10$\pm$0.13, [Si/H]=$-$1.18$\pm$0.13, [S/H]=$-$1.17$\pm$0.14 and [Fe/H]=$-$1.59$\pm$0.13. These values again indicate modest nucleosynthesis production and modest dust depletion.

The galaxy responsible for this absorber is not detected in our SINFONI data.

\subsection{Q1356$-$1101, \zabs=2.5009 (2.9668)}

This quasar was part of the Parkes radio survey and its absorber was first reported by Ellison et al. (2001) as part of the CORALS survey which aimed at looking for absorbers in radio-selected quasars. These authors derived a column density of \lognhi=20.40$\pm$0.49. Akerman et al. (2005) measure the metallicity by fitting at \zabs=2.5009:
 [Zn/H]$<-$1.32, [Cr/H]= $-$1.18, [Fe/H]=$-$1.25. Noterdaeme et al. (2008) derive a molecular fraction log f$<-$5.18 and report a $\Delta$ v=71 km/s for the metal lines.

Ellison et al. (2001) also reported an absorber associated with the quasar itself at \zabs=2.9668 with \lognhi=20.78$\pm$0.47. Akerman et al. (2005) analysed this absorber and derive: [Zn/H]$<-$1.48, [Cr/H]$<-$1.65, [Fe/H]=$-$1.54 and [Si/H]=$-$1.31. Noterdaeme et al. (2008) derive a molecular fraction log f$<-$6.75 and report a $\Delta$ v=30 km/s for the metal lines.

This latter object is not covered by our SINFONI data, while the former galaxy responsible for the prime absorber at \zabs=2.5009 is not detected.

\subsection{Q1454$+$1210, \zabs=2.2550 (2.4690, 3.1717)}

The absorption-line systems towards this quasar discovered by Hazard et al. (1986), have first been studied by Wolfe et al. (1986). The DLA at \zabs=2.4690 was reported by Lanzetta et al. (1991) as well as two other absorption systems at \zabs=2.2550 and 3.1717. 

Petitjean et al. (2000) studied the metal content of the DLA at \zabs=2.4690 with \lognhi=20.39$\pm$0.10. They found [Fe/H]=$-$2.54$\pm$0.12 and [Zn/H]=$-$1.95$\pm$0.16. Noterdaeme et al. (2008) state [Si/H]=$-$2.27$\pm$0.13 over $\Delta$ v = 35 km/s and log f$<-$4.50.

Dessauges-Zavadsky et al. (2003) have made a detailed study of the \zabs=2.2550 system and measure \lognhi=20.30$\pm$0.15. These authors fitted 5 components spread over $\sim$ 130 km/s. They obtained [Si/H]$>-$0.40, [Fe/H]=$-$1.47$\pm$0.17, [Ni/H]$<-$1.02, [Zn/H]=$-$1.12$\pm$0.19 and [Cr/H]=$-$1.11$\pm$0.20. We note that in this system, as for the absorber at \zabs=2.6147 towards Q2350$-$0052, [Si/H] is higher than [Zn/H], which is unusual. The \mgii\ doublet is also covered but is so highly saturated at the UVES resolution so that no useful lower limit can be deduced.

Dessauges-Zavadsky et al. (2003) also includes a detailed study of the \zabs=3.1717 system for which the \nhi\ is not covered. Petitjean et al. (2000) estimated the hydrogen column density of this system from the equivalent width measurement by Bechtold (1994) and give \lognhi=19.70. Dessauges-Zavadsky et al. (2003) fitted 2 components spread over $\sim$ 25 km/s. They obtained [O/H]$<-$1.73, [C/H]=$-$1.92$\pm$0.26, [Si/H]=$-$1.62$\pm$0.15, [Al/H]=$-$1.88$\pm$0.15, [Fe/H]=$-$1.87$\pm$0.16, [Zn/H]$<-$0.36 and [Cr/H]$<-$0.74 the last two values from non-detections. These indicate low nucleosynthesis products without any indication of dust.The \mgii\ doublet is not covered by the UVES spectrum.

Warren et al. (2001) searched for the galaxy counterpart to the main DLA at \zabs=2.4690 but do not find candidates in that field in their NICMOS search down to H$_{AB}$=25.0. Christensen et al. (2007) also studied that field to search the 3 absorbers with Integral Field Unit Potsdam Multi Aperture Spectrophotometer (PMAS) but report candidates that have not been subsequently followed-up for confirmation. Similarly, our SINFONI observations show no indication of a \ha\ emission at that redshift, nor at \zabs=2.2550. Note that our data do not cover the \ha\ emission line of the \zabs=3.1717 absorber.

\subsection{Q1631$+$1156, \zabs=0.9008 (0.5316, 1.3786)}

Junkkarinen, Hewitt \& Burbidge (1991) reported a \mgii\ absorber at \zabs=0.5316 and another one \zabs=0.9008. Aldcroft et al. (1994) mention a \civ\ absorber at \zabs=1.3786 and confirm the \mgii\ and \feii\ lines at the position reported by Junkkarinen, Hewitt \& Burbidge (1991) with EW(\mgii\ 2803)=1.35$\pm$0.07\AA\ and EW(\mgii\ 2803)=0.69$\pm$0.09\AA\ (see also Rao et al. 2006). The \hi\ column density of the \zabs=0.5316 was measured by Rao et al. (2003) to have \lognhi=20.70$^{+0.08}_{-0.10}$ from HST/STIS observations, while the one at \zabs=0.9008 has \lognhi=19.70$^{+0.03}_{-0.04}$. Rao et al. (2003) also obtained UBRJK images of the field and identified a disk galaxy at 3.2" away from the background quasar. They found that the observed Spectral Energy Distribution (SED) of the galaxy may be described by a young galaxy at \zem=0.5316 that had recently undergone a single burst of star formation. Follow-up spectroscopy confirms that this galaxy is at \zem=0.529 (Chen \& Lanzetta, 2003) and the corresponding projected distance is $\rho$=14.1 h$^{-1}$ kpc. The redshifted \ha\ line of this absorber is not covered by our SINFONI data, as is also the case for the \zabs=1.3786 system.

For the system at \zabs=0.9008, Ellison (2006) report [Zn/H]=$-$0.18 and [Fe/H]=$-$0.86. Meiring et al. (2008) used higher resolution data from the MIKE spectrograph to study this system and derive [Fe/H]=$-$1.06$\pm$0.06 from fitting 6 components 
 and [Zn/H]$<-$0.15 from non-detection of the lines. Rao et al. (2003) also searched for the \zabs=0.9008 system in their images but do not detect it. They derive a surface brightness limit of $\mu_K >$ 21.6 mag/arcsec$^{-2}$. These authors also consider that the quasar PSF was subtracted reasonably cleanly, since the PSF was well defined by several stars brighter than the quasar in each frame. 

The galaxy responsible for this absorber was the main target of our SINFONI search but its \ha\ line remains undetected.

\subsection{Q2059$-$0528, \zabs=2.2100}

This is another quasar discovered as part of the SDSS survey and reported in its data release 1 (Schneider et al. 2003). Herbert-Fort et al. (2006) have studied this system at \zabs=2.210 in detail and report [Si/H]=$-$1.00$\pm$0.30. Prochaska et al. (2007) list \lognhi=20.80$\pm$0.20 and [Zn/H]=$-$0.53$\pm$0.11, [Fe/H]=$-$1.30$\pm$0.11. The \mgii\ doublet is not covered by these data. 

The galaxy responsible for this absorber is not detected in our SINFONI data.

\subsection{Q2102$-$3553, \zabs=2.5070 (1.2430, 1.3999, 3.0826)}

This quasar was discovered by Warren, Hewett \& Osmer (1991). Dessauges-Zavadsky et al. (2003) studied its intervening sub-DLA system at \zabs=2.5070 from UVES spectroscopy and derive \lognhi=20.21$\pm$0.10. The metal-lines are well fitted by 2 components over $\Delta$ v = 20 km/s: [O/H]=$-$1.50$\pm$0.24, [C/H]=$-$1.86$\pm$0.22, [Si/H]=$-$1.04$\pm$0.15, [N/H]=$-$3.47$\pm$0.21, [Fe/H]=$-$2.24$\pm$0.10, [Zn/H]$<-$0.44$\pm$0.14 and [Cr/H]=$-$0.83$\pm$0.16. The \mgii\ doublet is not covered by this UVES spectrum.

Narayanan et al. (2007) report two weak \mgii\ doublets: one at \zabs=1.2430 with EW(\mgii\ 2796)=0.015$\pm$0.001\AA\ and EW(\mgii\ 2803)=0.010$\pm$0.002\AA\ as well as another at \zabs=1.3999 with EW(\mgii\ 2796)=0.109$\pm$0.002\AA\ and EW(\mgii\ 2803)=0.061$\pm$0.003\AA.

Finally, Petitjean et al. (2008) studied the molecular content of the DLA at \zabs=3.0826 and derive \lognhi=20.98$\pm$0.08. They also report 3-component fits leading to: [Fe/H] =$-$1.97$\pm$0.08, [O/H]=$-$1.58$\pm$0.09,  [N/H]=$-$2.86$\pm$0.08, [S/H]=$-$1.76$\pm$0.09 and [Si/H] =$-$1.66$\pm$0.09. Again, this indicates low nucleosynthetic production and little dust. The \mgii\ doublet is not covered by this UVES spectrum.

Pettini et al. (1995) first reported the detection of the \lya\ emission in the trough of the absorber at \zabs=3.0826 and derive F(\lya)=7$\times$10$^{-17}$ erg/s/cm$^2$ corresponding to a luminosity L(\lya)=5$\times$10$^{42}$ erg/s with a shift of $+$470 km/s. Leibundgut \& Robertson (1999) confirm this \lya\ feature described by two components separated by 5 \AA\ and found the shift to be of the order $+$490 km/s. Nevertheless, the \ha\ line of that prime target is not detected in our SINFONI data leading to F(\ha)$<$1.2$\times$10$^{-17}$ erg/s/cm$^2$, corresponding to SFR$<$2.8 M$_\odot$/yr. These results indicate a ratio \lya/\ha$>$6, i.e. consistent with the \lya\ case B recombination rate.

The other absorbers are not covered by our observations.

\subsection{Q2222$-$0946, \zabs=2.3541 (2.8647)}

A \civ\ absorber is reported at \zabs=2.8647 from these data (York et al., private communication) but the emission lines of the \civ\ absorber at \zabs=2.8647 are not covered by our observations.

\subsection{Q2313$-$3704, \zabs=2.1821 (0.3399)}

Again, this quasar was part of the Parkes radio survey and its absorber was first reported by Ellison et al. (2001) as part of the CORALS survey. These authors derived a column density of \lognhi=20.48$\pm$0.10. Akerman et al. (2005) measure the metallicity by fitting at \zabs=2.1821:
 [Zn/H]$<-$1.29, [Cr/H]$<-$1.60, [Fe/H]=$-$1.70 and [Si/H]=$-$1.52. Noterdaeme et al. (2008) derive a molecular fraction log f$<-$5.00 and report a $\Delta$ v=77 km/s for the metal lines.

Mshar et al. (2007) report another \mgii\ absorber with associated \feii\ at \zabs=0.3399. They report N(\feii)$>$14.41 and N(\mgii)$>$14.07.

The galaxy responsible for the main absorber at \zabs=2.1821 is not detected in our SINFONI data, while the \mgii\ system at \zabs=0.3399 is not covered by our data at the \ha\ position.

\subsection{Q2350$-$0052, \zabs=2.6147 (0.8629, 1.0793, 2.0470, 2.4263, 2.9298)}

There are two DLAs at \zabs = 2.4263 and \zabs = 2.6147 toward Q2350$-$0052. These DLAs were discovered as strong \lya\ absorptions in a low spectral-resolution ($\sim$5\AA) spectrum by Sargent et al. (1989). 

Prochaska et al. (2001) studied the system at \zabs=2.4263 with HIRES spectra and derive \lognhi=20.50$\pm$0.10, [Al/H]$>-$1.051, [Si/H]=$-$0.695$\pm$0.102, [Cr/H]$<-$1.457, [Fe/H]=$-$1.386$\pm$0.101 and [Ni/H]=$-$1.400$\pm$0.144. Noterdaeme et al. (2007) further analysed the absorber with UVES spectra and using 14 components derive a molecular hydrogen fraction log f=$-$1.69. The Si and Fe imply a system with 1/5$^{th}$ solar metallicity in the gas phase and grain depletion. Depleted gas phase elements normally have a pattern with Al much more depleted than Si, so in this regard, the [Al/H] is anomalous, but the presence of dust would also explain the H$_2$.

For the system at \zabs=2.6147, Prochaska et al. (2001) derive \lognhi=21.30$\pm$0.10, [Al/H]$>-$2.651, [Si/H]=$-$1.968$\pm$0.123, [Cr/H]=$-$2.296$\pm$0.117, [Fe/H]=$-$2.227$\pm$0.133, [Ni/H]=$-$2.357$\pm$0.124 and [Zn/H]$<-$2.099. No molecular hydrogen is detected at this redshift down to log f $<-$7.2 according to Noterdaeme et al. (2007).

Lanzetta et al. (1991) also report an absorber with unknown \nhi\ at \zabs=2.0470. An additional LLS was reported at \zabs=2.9298 (Stengler-Larrea et al. 1995) with associated \civ\ features (Lu \& Savage 1993). Finally, Khare, York \& Green (1989) report \mgii\ absorbers at \zabs=0.8629 and \zabs=1.0793.

Bunker et al. (1999) use long-slit K-band spectroscopic on UKIRT to try to detect the galaxies responsible for these two absorbers. They did not find any \ha\ emission line leading to upper limits of F(\ha)$<$7.53$\times$10$^{-17}$ erg/s/cm$^2$, i.e. SFR$<$7.18 h$^{-2}$M$_{\odot}$/yr at \zabs=2.4263 and F(\ha)$<$10.5$\times$10$^{-17}$ erg/s/cm$^2$, i.e. SFR$<$11.9 h$^{-2}$M$_{\odot}$/yr at \zabs=2.6147. Similarly, we do not detect any of these two absorbers in our SINFONI data at limits that are lower than those set by Bunker et al. (1999). For \zabs=2.6147, we derive F(\ha)$<$1.5$\times$10$^{-17}$ erg/s/cm$^2$, i.e. SFR$<$3.8 M$_{\odot}$/yr.

For the non-prime targets, only the \zabs=2.0470 and \zabs=2.4263 are covered by our SINFONI data but they remain undetected.

\subsection{Q2352$-$0028, \zabs=1.0318 (0.8730, 1.2468)}

For the \zabs=0.8730, Rao, Turnshek \& Nestor (2006) measure \lognhi=19.18$^{+0.08}_{-0.10}$. The corresponding EW(\mgii\ 2803)=0.816$\pm$0.075\AA. Meiring et al. (2009) used 10 components to fit the absorber over $\Delta$ v $\sim$ 225 km/s and derive: [Zn/H]$<-$0.14, [Fe/H]=$-$1.17$\pm$0.09.

For the \zabs=1.2468, Rao, Turnshek \& Nestor (2006) measure \lognhi=19.60$^{+0.18}_{-0.30}$. The corresponding EW(\mgii\ 2803)=2.261$\pm$0.120\AA. Meiring et al. (2009) used 15 components to fit the absorber over $\Delta$ v $\sim$ 410 km/s and derive: [Zn/H]$<-$0.70 and [Fe/H] =$-$0.86$\pm$0.24

In our SINFONI observations, we do not detect the galaxy responsible for the absorber at \zabs=0.8730 and the absorber at \zabs=1.2468 is not covered by our data.

\bsp

\label{lastpage}

\end{document}